\documentclass[aps,twocolumn,showpacs,amsmath,amssymb,amsfonts]{revtex4}
\usepackage{graphicx}
\usepackage{bm}
\usepackage{color}

\newcommand{\tst}{\textstyle}

\newcommand{\mbf}{\mathbf}
\newcommand{\mrm}{\mathrm}

\begin{document}

\author{Zbigniew Idziaszek}
\affiliation{CNR-INFM BEC Center, I-38050 Povo (TN), Italy \\and Center for Theoretical Physics, Polish Academy of Sciences, 02-668 Warsaw, Poland}
\author{Tommaso Calarco}
\affiliation{CNR-INFM BEC Center and ECT*, I-38050 Povo (TN), Italy\\
and ITAMP, Harvard University, Cambridge, MA 02138, USA}
\author{Peter Zoller}
\affiliation{Institute for Quantum Optics and Quantum Information of the Austrian
Academy of Sciences\\
and Institute for Theoretical Physics, University of Innsbruck, A-6020
Innsbruck, Austria}

\title{Controlled collisions of a single atom and ion guided by movable trapping potentials}

\begin{abstract}
We consider a system composed of a trapped atom and a trapped ion. The ion charge induces in the atom an electric dipole moment, which attracts it with an $r^{-4}$ dependence at large distances. In the regime considered here, the characteristic range of the atom-ion interaction is comparable or larger than the characteristic size of the trapping potential, which excludes the application of the contact pseudopotential. The short-range part of the interaction is described in the framework of quantum-defect theory, by introducing some short-range parameters, which can be related to the $s$-wave scattering length. When the separation between traps is changed we observe trap-induced shape resonances between molecular bound states and vibrational states of the external trapping potential. Our analysis is extended to quasi-one-dimensional geometries, when the scattering exhibit confinement-induced resonances, similar to the ones studied before for short-range interactions. For quasi-one-dimensional systems we investigate the effects of coupling between the center of mass and relative motion, which occurs for different trapping frequencies of atom and ion traps. Finally, we show how the two types of resonances can be employed for quantum state control and spectroscopy of atom-ion molecules.
\end{abstract}


\maketitle

\section{introduction}

Techniques developed in atomic physics during the last two decades allow the preparation of single atoms and ions in the laboratory. Single neutral atoms can be stored and laser cooled in far off-resonance laser traps (FORT)
\cite{Grangier,Meschede}, and arrays of atoms can be prepared in an optical lattice via a Mott insulator phase, where exactly one atom is stored per lattice site \cite{Bloch}. Similar techniques are being pursued in the context of atom chips \cite{microtraps}. Furthermore, single ions and arrays of ions can be stored in Paul and Penning traps, and sideband laser cooling allows to prepare ions in the vibrational ground state of these trapping potentials \cite{Leibfried}.   The controlled preparation, manipulation and measurement of electronic internal and motional states of laser driven single atoms and ions, combined with the possibility of controlled entanglement of atoms and ions provides the basic ingredients to for investigations of fundamental aspects of quantum mechanics, and applications such as high-precision measurements and quantum information.  In quantum information processing entanglement of qubits stored in internal states of  single atoms or ions is achieved by controlled interactions between particles, either in the form of switchable qubit-dependent two-particle interactions or via an auxiliary collective mode of the systems, which serves as a quantum data bus. In particular, for neutral atoms  cold controlled collisions of atoms stored in movable spin-(i.e.~qubit-)dependent optical lattices has been proposed as a means to entangle atomic pairs, and has experimentally implemented to generate  $N$-atom cluster states \cite{Bloch2}. 

\begin{figure}
   \includegraphics[width=8.6cm,clip]{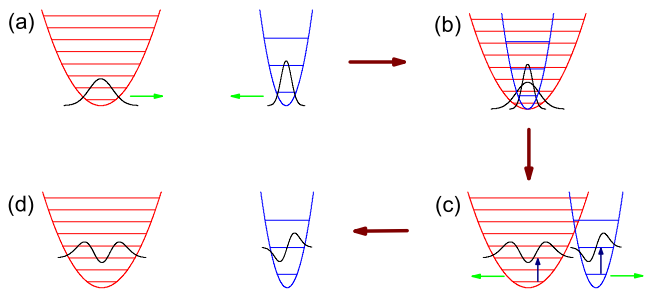}
     \caption{
     (Color online) Schematic drawing of a controlled collision between  a single atom and a single ion, whose center-of-mass wavepackets are guided by time-dependent atom trap and ion trap, respectively. Labels distinguish different phases of the process: a) initially particles prepared in the motional ground state; b) collision (overlap of the wave packets); c) excitation of the motional states in the traps; d) particles in some excited states after the collision.
     \label{Fig:ContrColl}
   }
\end{figure}

In the present work we study the  {\em controlled cold collision of a single atom and a single ion}, where the atom (ion) is prepared in a given motional state of an atom (ion) trap, and we move the traps to guide the atom and ion wave packet to ``collide'' for a given time (see Fig.~\ref{Fig:ContrColl}). Our focus is the development of a {\em quantum-defect} formalism for trapped atoms and ions, in a form which is convenient for future applications and extensions, in particular in the context of quantum information processing to swap qubits stored in atoms and ions, and the entanglement of these qubits in a controlled collision. The atom-ion collision is governed by the potential  $V(r) \rightarrow -\alpha e^2/(2r^4)$ ($r \rightarrow \infty$), where $\alpha$ is the dipolar polarizability -- to be contrasted to a van der Waals potential  $V(r) \rightarrow -C_{6}/(r^6)$ which represents the collisional interactions between e.g. alkali atoms in their electronic ground state. Below we will calculate the dynamics of the interacting atom-ion system for a given time-dependence of the motion of the trap. This includes the transition to excited trap states after the collision, the possible formation of an excited atom-ion (molecule) complex, and the description of trap induced resonances.

%%%%%%%%%%%%%%%%%% Figure 1 %%%%%%%%%%%%%%%%%%%%%%
%%%%%%%%%%%%%%%%%%%%%%%%%%%%%%%%%%%%%%%%%%%%%%%%%%%%%

To study the controlled atom-ion collisions we first derive an effective Hamiltonian, describing effective net forces acting on the particles moving in rapidly changing laser and rf fields. In our approach, we include in the Hamiltonian only the asymptotic part of the atom-ion potential, whereas the short-range interactions are taken into account by imposing appropriate boundary conditions on the wave function at $r \rightarrow 0$. In the regime of cold collisions considered here, the boundary conditions can be expressed in terms of a single quantum-defect parameter, independent of the collisional {\em energy} and {\em angular momentum}. The quantum-defect approach is dictated by the relatively long-range character of the atom-ion interaction, which exceeds the typical size of the trapping potentials, and, in contrast to the neutral atoms, excludes the applicability of the pseudopotential.

Our method to describe the dynamics of the controlled collision is based on the application of the correlation diagrams, i.e. energy spectra as a function of the trap separation. Such correlation diagrams, widely used in quantum chemistry to characterize reactions of diatomic molecules, connect in our case the asymptotic vibrational states with the molecular and vibrational states at zero trap separations. At the intermediate distances the energy curves exhibit avoided or diabatic crossings, depending on the symmetry of eigenstates and of the coupling term in the Hamiltonian. In the atom-ion system the avoided crossing can be attributed to the resonances between molecular and vibrational states, that appear when the energy of a vibrational level, coincide with the energy of a molecular state shifted by the external trapping. The dynamics in the vicinity of such avoided crossings can be accurately described in the framework of the Landau-Zener theory. In this paper we determine the level splitting at the avoided crossing, at different trap separations, and for different symmetries of the molecular states. In this way we can characterize the time scale appropriate for adiabatic or diabatic traversing of a given avoided crossing.

In our paper we consider two different geometries of the trapping potentials: spherically symmetric traps and very elongated cigar-shape traps (quasi-1D traps). The former case requires full three-dimensional treatment, while the latter one can be described using an effective one-dimensional Hamiltonian. In the latter case, the tight transverse confinement effectively renormalizes the one-dimensional quantum-defect parameters. In addition, very elongated cigar-shape traps exhibit another type of resonances, that can be observed at zero-trap separations, but for changing ratio of the transverse confinement to the s-wave scattering length of the atom-ion interaction. Such confinement-induced resonances appear when the energy of the colliding particles coincide with energy of a bound state lifted up by the tight transverse potential.

As already outlined above, the counterpart to the idea of ultracold controlled trap-guided collisions has been implemented in optical lattices, where application of spin-dependent potentials has allowed to entangle atoms between neighboring sites \cite{Bloch2}. Moreover, resonance phenomena in trapped systems, similar to the ones considered here, have been already thoroughly investigated for interactions between neutral atoms. In quasi-1D systems, collisions between neutral atoms exhibit confinement-induced resonances \cite{Olshanii}, whereas displacement of the trapping potentials leads to trap-induced resonances \cite{Stock}.

The paper is organized as follows. In section~\ref{Sec:BasicSetup} we present our model that we use to describe the controlled atom-ion collisions. In particular, in section~\ref{Sec:EffHam} we introduce an effective Hamiltonian of a single trapped atom and ion. Section~\ref{Sec:QuantDef} discuss the quantum-defect treatment of the short-range part of the interaction potential.
In sections~\ref{Sec:Symm3D} and \ref{Sec:Quasi1D} we discuss our model for two particular geometries of the trapping potentials, spherically symmetric traps of the same frequency and elongated cigar-shape traps with the same transverse trapping frequency, respectively. We extend the quantum-defect treatment to quasi-1D systems in section~\ref{Sec:QuantDef1D}. Our approach to describe the dynamics of the atom-ion system is discussed in section~\ref{Sec:TimeDep}. Section~\ref{Sec:Results} presents the results of our calculations. We start from discussing the properties of the scattering in quasi-1D systems in the presence of long-range $r^{-4}$ potential, in particular, the problem of determining of 1D quantum-defect parameters.
Sections~\ref{Sec:Rel1D} and \ref{Sec:Rel3D} present adiabatic energy curves and adiabatic eigenstates in, respectively, 1D and 3D traps with the same trapping frequencies for the atom and the ion. The dynamics in the vicinity of the avoided crossing is discussed in section~\ref{Sec:AvoidedCross}, where we apply the Landau-Zener theory and semiclassical methods to address this problem. Section~\ref{Sec:Full1D} discuss most complicated case of different trapping frequencies, focusing on the 1D atom-ion motion. The outlook and final conclusions are presented in section~\ref{Sec:Outlook}. Finally appendix~\ref{App:Micro} presents the microscopic derivation of the effective Hamiltonian and appendix~\ref{App:Pseudo} describes the details of the pseudopotential treatment of the scattering in quasi-1D traps.

\section{Basic setup and model}
\label{Sec:BasicSetup}

\begin{figure}
   \includegraphics[width=8cm,clip]{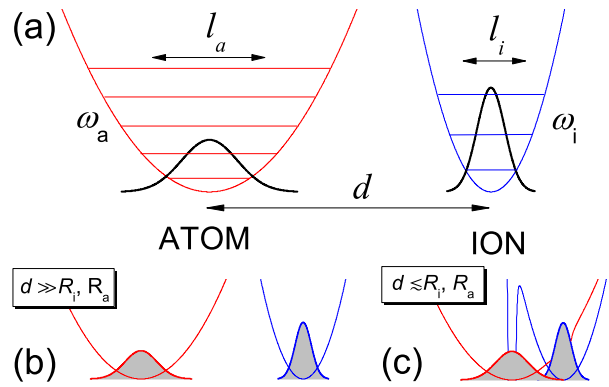}
     \caption{ 
     (Color online) Panel (a): Schematic drawing of the considered setup consisting of a trapped atom and ion. Panels (b) and (c): Long and short-range regimes, respectively, in the controlled collisions between a trapped atom and ion.
     \label{Fig:Setup}
   }
\end{figure}

We consider a system consisting of a single atom and single ion, stored in their respective trapping potentials. Single atoms can be trapped in experiments with optical tweezers \cite{Grangier,Meschede} or optical lattices \cite{Bloch}. Such potentials are created in far-detuned laser fields due to the AC Stark effect. In another experimental technique, on atom chips \cite{microtraps}, single atoms can be trapped in microtraps created by electric and magnetic fields around wires and electrodes. On the other hand, single ions can be confined in radio-frequency traps, which use a rapidly oscillating electric field to create an adiabatic trapping potential for the ion charge \cite{Leibfried}. Typically, the trapping potentials close to the trap center are with a good approximation harmonic, and in our approach we consider the simple picture of the atom and ion confined in the harmonic potentials with frequencies $\omega_a$ and $\omega_i$, respectively, with $d$ denoting the displacement between the traps (c.f. Fig.~\ref{Fig:Setup}.a). We assume that the atom and ion are prepared initially in a given trap state, e.g. the ground state of the potential, and that the atom and ion are brought to collision by overlapping their center-of-mass (COM) wave packets, i.e. we change $d(t)$ as a function of time as outlined in Fig.~\ref{Fig:ContrColl}.

In studying this collision with COM wave packets guided by movable trapping potentials we will study two cases of (i) three-dimensional (3D) dynamics, where both traps are assumed to be spherically symmetric, and (ii) the case of quasi-1D traps, i.e. elongated, cigar-shaped traps which can be described as an effectively 1D collision dynamics. This second case is conceptually simpler, but also provides collisional features specific to 1D situation  (e.g. the appearance of 1D confinement-induced resonances, reminiscent of resonances in \cite{Olshanii}).  The values of trapping frequencies in atom and ion traps  are typically quite different, thus it is natural to consider $\omega_a \neq \omega_i$. For such condition, however, the COM and relative motion are coupled, which significantly complicates the theoretical description. Apart from this general regime, we consider also the special case of $\omega_a = \omega_i$, when COM and relative motions can be separated. In this case we perform the full diagonalization for the 3D problem, which is numerically very demanding in the general case of $\omega_a \neq \omega_i$.

%%%%%%%%%%%%%%%%%% Figure 2 %%%%%%%%%%%%%%%%%%%%%%
%%%%%%%%%%%%%%%%%%%%%%%%%%%%%%%%%%%%%%%%%%%%%%%%%%%%%

At large distances, the interaction between atom and ion is given by
a potential $V(r) \sim -\alpha e^2/(2r^4)$, where $\alpha$ is the
dipolar polarizability. This potential originates from the
attraction between the ion charge and the electric dipole that it
induces on the atom \footnote{For atomic states with a permanent
quadrupole moment, the long-range part of the potential comes from
the interaction of ion charge with the atom quadrupole moment.}. At
short distances the interaction is more complicated; however, it
turns out that a detailed knowledge of the form of core region of
the potential is not necessary, and that it can be subsumed in our
model by introducing a set of (energy independent) quantum-defect
parameters.

Let us focus now on characteristic scales that can be associated with the atom-ion interaction. For the sake of clarity table~\ref{Tab1} summarizes all the length scales that are introduced in the course of the paper. We define its characteristic length as $R^\ast \equiv \sqrt{ \alpha e^2 \mu /\hbar^2}$, and characteristic energy as $E^{\ast} = \hbar^2/\left(2 \mu (R^\ast)^2\right)$, where $\mu$ is the reduced mass. Other characteristic lengths and energies are given by the ground states of the trapping potentials. We associate to the trapping frequencies the harmonic oscillator lengths for atom (ion): $l_a = \sqrt{\hbar/(m_a \omega_a)}$ ($l_i = \sqrt{\hbar/(m_i \omega_i)}$). For optical and rf traps, the trapping frequencies are typically $\omega_a = 2 \pi \times 10-100 \textrm{kHz}$ and $\omega_i = 2 \pi \times 0.1 - 10 \textrm{MHz}$. In table~\ref{Tab} we collected some example values of $R^\ast$, $l_i$, $l_a$, and $E^{\ast}$ for a few systems of alkali atoms and alkaline earth ions. One can observe that the range of the polarization potential is comparable or larger than the size of the trapped ground state. Therefore, for the trapped particles the polarization interaction cannot be replaced with the standard contact pseudopotential familiar from $s$-wave atom-atom scattering in the context, e.g., of Bose-Einstein condensation.

%%%%%%%%%%%%%%%%%% Table 1 %%%%%%%%%%%%%%%%%%%%%%
\begin{table}
\begin{ruledtabular}
\begin{tabular}{lp{5.5cm}}
Definition & Description \\
\hline
$R^{\ast} = \sqrt{\alpha e^2 \mu/\hbar^2}$ & characteristic length of the polarization potential \\
$l_i = \sqrt{\hbar/(m_i \omega_i)}$ & harmonic oscillator length of ion trap\\
$l_a = \sqrt{\hbar/(m_a \omega_a)}$ & harmonic oscillator length of atom trap\\
$l = \sqrt{\hbar/(\mu \omega)}$ & harmonic oscillator length for relative degrees of freedom (axial direction in case of quasi 1D traps)\\
$l_\perp = \sqrt{\hbar/(\mu \omega_\perp)}$ & harmonic oscillator length in the transverse direction for relative degrees of freedom \\
${\displaystyle R_i = \left(\frac{\alpha e^2}{m_i \omega_i^2}\right)^{1/6}}$
& distance at which the polarization potential becomes equal to the ion trappping potential\\
${\displaystyle R_a = \left(\frac{\alpha e^2}{m_a \omega_a^2}\right)^{1/6}}$
& distance at which the polarization potential becomes equal to the atom trappping potential \\
${\displaystyle R_\mrm{rel} = \left(\frac{\alpha e^2}{\mu \omega^2}\right)^{1/6}}$ & distance at which the polarization potential becomes equal to the trappping potential for relative degrees of freedom \\
${\displaystyle R_\perp = \left(\frac{\alpha e^2}{\mu \omega_\perp^2}\right)^{1/6}}$ & distance at which the polarization potential becomes equal to the trapping potential in the transverse direction for relative degrees of freedom \\
$R_\mrm{1D} = \max (R_\perp,l_\perp)$ & boundary of the quasi-1D regime \\
$R_0$ & size of the core region of the atom-ion complex 
\end{tabular}
\end{ruledtabular}
\caption{\label{Tab1} Definitions of the length scales used throughout the paper.
}
\end{table}
%%%%%%%%%%%%%%%%%%%%%%%%%%%%%%%%%%%%%%%%%%%%%%

%%%%%%%%%%%%%%%%%%%%%%% Table 2 %%%%%%%%%%%%%%%%%%%%%%
\begin{table}
\begin{ruledtabular}
\begin{tabular}{lllll}
& $R^{\ast}$($a_0$) & $l_i$($a_0$) & $l_a$($a_0$) & $E^{\ast}/h$ (kHz) \\
\hline
$^{40}$Ca$^{+}$ + $^{87}$Rb & 3989 & 300 & 644 & 4.143 \\
$^{9}$Be$^{+}$ + $^{87}$Rb & 2179 & 632 & 644 & 46.59   \\
$^{40}$Ca$^{+}$ + $^{23}$Na & 2081 & 300 & 1252 & 28.56   \\
\end{tabular}
\end{ruledtabular}
\caption{\label{Tab} Characteristic distance $R^{\ast}$, characteristic energy $E^{\ast}$, harmonic oscillator length $l_i$ for ion trap of $\omega_i = 1$MHz, and harmonic oscillator length $l_a$ for atom trap of $\omega_a = 100$kHz, for some combinations of alkali earth ions and alkali atoms.
}
\end{table}
%%%%%%%%%%%%%%%%%%%%%%%%%%%%%%%%%%%%%%%%%%%%%%%%%%%

In the interactions between trapped atom and ion, one can distinguish two different regimes depending on the relative distance $d$ between traps: (i) $d \gg R_{i}, R_{a}$, (ii) $d \lesssim R_{i} \sim R_{a}$, where $R_\nu = \left(\alpha e^2 /m_\nu \omega_\nu^2\right)^{1/6}$ for $\nu=i,a$ is some characteristic distance at which the atom-ion interaction becomes comparable to the trapping potential of the ion and atom, respectively. Because of the weak dependence of $R_\nu$ on $\omega_\nu$ and $m_\nu$, the characteristic distances $R_i$ and $R_a$ are roughly the same. The two regimes (i) and (ii) correspond to the cases where the motion in the traps is weakly or strongly affected by the atom-ion interactions, respectively. The discussed regimes are illustrated schematically in panels (b) and (c) of Fig.~\ref{Fig:Setup}, and they correspond to snap shots in the collision process  when the wavepackets of the particles do not overlap / do  overlap during the controlled collision. In the regime (i) of large distances, the atom-ion interaction can be treated perturbatively as a distortion of the trapping potential. In this limit the system can be described in terms of two coupled harmonic oscillators, and all the dynamics can be solved analytically. We discuss this case in more details in section~\ref{Sec:Large}. On the other hand, in the regime (ii) the description is more difficult, since it requires inclusion of the short-range part of the interaction, and full treatment of the long-range $r^{-4}$ part. In the present paper we focus mainly on the latter case.

\subsection{Effective Hamiltonian}
\label{Sec:EffHam}

We adopt the following time-dependent Hamiltonian to describe the system of a single trapped atom and a single trapped ion (c.f. Fig.~\ref{Fig:Setup}):
\begin{align}
\label{H3D}
H(t) = & \sum_{\nu = i,a} \left[ \frac{\mbf{p}_\nu^2}{2 m_\nu} +
\frac{1}{2} m_\nu \omega_\nu^2 (z_\nu - d_\nu(t))^2
+ \frac{1}{2} m_\nu \omega_{\perp \nu}^2 \rho_\nu^2
\right]\nonumber \\
& + V(|\mbf{r}_i-\mbf{r}_a|),
\end{align}
Here the label $i$ ($a$) refers to the ion (atom) respectively, $\mbf{p}$ and $\mbf{r}$ are the momentum and position operators, $d_\nu(t)$ denotes the positions of the atom and ion traps, respectively, that can be controlled in the course of dynamics, and $\rho^2 = x^2 + z^2$. We assume that the trapping potentials are axially symmetric and displaced along the axis of symmetry. The trapping frequencies are denoted by $\omega_\nu$ and $\omega_{\perp \nu}$ for the axial and transverse directions, respectively. Finally, $V(r)$ denotes the interaction potential between the atom and the ion. At large distances the main contribution to this interaction comes from the polarization of the atomic cloud, $V(r) \sim -\alpha e^2/(2r^4)$.

A microscopic derivation of the Hamiltonian \eqref{H3D} is presented in appendix~\ref{App:Micro}. The basic idea of this derivation is based on the application of the Born-Oppenheimer approximation to the motion of the outer shell electrons, and followed by a time averaging over fast time scale of the rf and laser frequencies. In this picture, the Hamiltonian \eqref{H3D} represents effective net interactions felt by particles moving in the rapidly changing time-dependent potentials. The interaction potential $V(r)$ can be identified with the adiabatic Born-Oppenheimer curve of the electronic ground-state, that depends on the difference between the atom and ion COM coordinates. 

\subsection{Quantum-defect theory}
\label{Sec:QuantDef}

We denote by $R_0$ the characteristic distance at which $V(r)$ starts to deviate from the asymptotic $r^{-4}$ law. Typically, $R_0$ is the size of the core region of the ion-atom complex, and is much smaller than all the other length scales in our problem. Therefore, we can describe the short-range part of interaction in the spirit of the quantum-defect theory. For $R_0 \ll R^{\ast}$, the parameters describing the short-range potential become {\em independent} of {\em energy} and {\em angular momentum}, since for $r \ll R^{\ast}$ the interaction potential is much larger than typical kinetic energies and heights of the angular-momentum barrier. This feature is a key ingredient of quantum-defect theory, where the complicated short-range dynamics can be  summarized in terms of few, energy-independent constants (phase shifts or quantum defects).

To analyze the short-distance behavior of the wave functions, we omit for the moment the trapping potentials in the Hamiltonian \eqref{H3D}, and consider only the part describing the scattering of atom from the ion in free space. In this case, the relative and COM degrees of freedom are decoupled, and the relative motion is governed by the Hamiltonian $H_0 = \mbf{p}^2/2 \mu + V(r)$. We apply the partial wave expansion to the relative wave function: $\Psi_\mrm{rel}(\mbf{r}) = \sum_l R_l(r) Y_{lm}(\theta,\phi)$ with $l$ denoting the angular momentum and $Y_{lm}(\theta,\phi)$ the spherical harmonics. For $r \gg R_0$, we set $V(r) = -\alpha e^2/(2r^4)$, which allows to solve radial Schr\"odinger equation in terms of the Mathieu functions of the imaginary argument \cite{Vogt,OMalley,Spector}. In this way we obtain the following short-distance behavior of the radial wave functions
\begin{equation}
\label{Sol1}
R_l(r,k) \sim \sin\left(R^\ast/r + \varphi_l(k)\right), \qquad r \ll \sqrt{R^{\ast}/k},
\end{equation}
where $\hbar k$ is the relative momentum, and $\varphi_l(k)$ are some short-range phases. As one can easily verify, the asymptotic solution 
\eqref{Sol1} fulfills the radial Schr\"odinger equation with the energy $\hbar^2 k^2/(2 \mu)$ and centrifugal barrier $\hbar^2 l(l+1)/(2 \mu r^2)$ terms neglected. The short-range phases constitute our quantum-defect parameters. For $R_0 \ll R^{\ast}$ we can assume that $\varphi_l(k)$ are independent of the energy and angular momentum: $\varphi_l(k) \equiv \varphi$, which reduces description of the short-range interaction to the single quantum-defect parameter $\phi$. In the calculations we replace $V(r)$ by its asymptotic $r^{-4}$ behavior, effectively letting $R_0 \rightarrow 0$ and imposing boundary condition stated by \eqref{Sol1} with the short-range phase $\phi$.

For $k=0$ the solution \eqref{Sol1} becomes valid at all distances. Utilizing the fact that zero-energy solution behaves asymptotically as
$\Psi(r) \sim 1 - a_s/r$ ($r \rightarrow \infty$), we can relate the short-range phase to the $s$-wave scattering length $a_s$
\begin{equation}
a_s = - R^{\ast} \cot \varphi
\end{equation}
In this way the knowledge of the scattering length $a_s$ allows us to calculate the quantum defect parameter $\varphi$.

\subsection{Symmetric 3D trapping potential and identical trapping frequencies}
\label{Sec:Symm3D}

In this section we consider 3D dynamics assuming, spherically symmetric trapping potentials $\omega_\nu= \omega_{\perp \nu}$ for $\nu = i,a$ and identical trapping frequencies for atom and ion: $\omega_i = \omega_a = \omega$. For such conditions the relative and CM motions are decoupled.

The COM motion is governed by the Hamiltonian of a harmonic oscillator with mass $M = m_i + m_a$ and frequency $\omega$, while the Hamiltonian of the relative motion is given by
\begin{align}
\label{Hrel}
H_\mrm{rel} = \frac{\mbf{p}^2}{2 \mu} +
\frac{1}{2} \mu \omega^2 (\mbf{r} - \mbf{d})^2 - \frac{\alpha e^2}{2 r^4},
\end{align}
where $\mbf{p}$ and $\mbf{r}$ are the relative-motion momentum and position operators, respectively, and  $\mbf{d} = \mbf{d}_a - \mbf{d}_i$. In the case of relative motion we define
\begin{equation}
\label{Rrel}
R_\mrm{rel} = \left(\alpha e^2 /\mu \omega^2\right)^{1/6} = (R^{\ast})^{1/3}
l^{2/3}
\end{equation}
as a characteristic length at which the atom-ion interaction is comparable to the trapping potential for the relative degrees of freedom, where $l = \sqrt{\hbar /(\mu \omega)}$ denotes harmonic oscillator length.

\begin{figure}
   \includegraphics[width=8.6cm,clip]{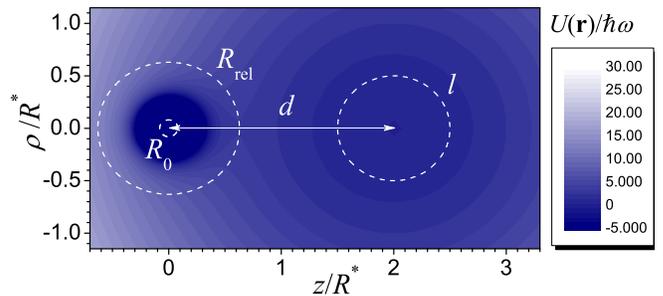}
     \caption{
     (Color online) Contour plot of the total potential $U(\mbf{r}) = V(\mbf{r}) +
     \frac{1}{2} \mu \omega^2 (\mbf{r} - \mbf{d})^2$ for the relative motion
     in the atom-ion system. The figure shows the case of spherically symmetric trapping potentials with the same trapping frequencies, separated by $\mbf{d} = (0,0,2 R^{\ast})$, and for $R^{\ast} = 2 l$ ($l = \sqrt{\hbar /(\mu \omega)}$).
     \label{Fig:Vrel}
   }
\end{figure}

Fig.~\ref{Fig:Vrel} illustrates a typical potential for the atom-ion relative motion.  In addition this figure depicts different lengths scales characteristic of our problem. In the presented case $R^{\ast} = 2 l$, and the distance between trap centers is $d=2 R^{\ast}$.

%%%%%%%%%%%%%%%%%% Figure 3 %%%%%%%%%%%%%%%%%%%%%%
%%%%%%%%%%%%%%%%%%%%%%%%%%%%%%%%%%%%%%%%%%%%%%%%%%%%%

\subsection{Quasi-1D trapping potentials}
\label{Sec:Quasi1D}

As a second geometry we consider cigar-shape traps with a transverse trapping frequency much larger than the axial one: $\omega_{\perp \nu} \gg \omega_\nu$ (quasi-1D traps) for $\nu = i,a$. For energies smaller than the excitation energy in the transverse direction, the motion in the transverse direction is frozen to zero-point oscillations, and the dynamics takes place along the weakly confined direction. Nevertheless, the transverse motion plays an important role at short distances, effectively renormalizing the short-range phase, as we will show in the next section. In this way quasi-1D traps offer the additional possibility of tuning the interactions, similarly to the case of neutral atoms exhibiting confinement-induced resonances \cite{Olshanii}.

Here, we consider only some particular situation when $\omega_{\perp i}=\omega_{\perp a}$ is the same for atom and ion. This simplifies our description of the renormalization effects, since in this case the transverse COM and relative motions can be separated. We expect that our results are also qualitatively valid in the general case of different transverse trapping frequencies.

To obtain an effective 1D Hamiltonian, describing the evolution of the wave packets along the $z$ axis, we decompose the total wave function of the atom and ion into a series over eigenmodes of the transverse part of the Hamiltonian. For total energies $E < 3 \hbar \omega_\perp$ only the 
transverse ground-state mode contributes to the total wave function
at large distances. In addition, condition $|z_i-z_a| \gg R_\perp$ assures 
that the axial and the transverse motions are decoupled and the wave function can be written as a product of the axial and the transverse components. Here, $R_\perp = \left(\alpha e^2 /\mu \omega_\perp^2\right)^{1/6}$ is some characteristic distance at which the atom-ion interaction becomes comparable to the transverse trapping potential, and the condition $|z_i-z_a| \gg R_\perp$ describes the regime where the transverse oscillation frequency is weakly modified by the atom-ion interaction. Hence, we have $\Psi(\mbf{r}_i,\mbf{r}_a) \rightarrow \psi_0(\rho_i,\rho_a) \Psi_\mrm{1D}(z_i,z_a)$ ($|z_i -z_a| \rightarrow \infty$) for $E < 3 \hbar \omega_\perp$. Here, $\psi_0$ is the ground state of the transverse part of the Hamiltonian: $\psi_0(\rho_i,\rho_a) = e^{- \omega_\perp  (m_i \rho_i^2+ m_a \rho_a^2)/2\hbar}/\pi^{1/2}$, and $\Psi_{1D}(z_i,z_a)$ denotes the axial part of the wave function. Substituting the decomposition over transverse modes into the Schr\"odinger equation with the Hamiltonian \eqref{H3D}, and retaining only the lowest transverse mode we obtain an effective 1D Hamiltonian, that governs the dynamics of $\Psi_\mrm{1D}(z_i,z_a)$
\begin{align}
\label{H1D}
H_\mrm{1D} = & \sum_{\nu = i,a} \left[ \frac{p_\nu^2}{2 m_\nu} +
\frac{1}{2} m_\nu \omega_\nu^2
(z_\nu - d_\nu)^2 \right]+ V_\mrm{1D}(|z_i-z_a|).
\end{align}
Here $V_\mrm{1D}(|z|)$ is the 1D interaction potential obtained by integrating out the transverse degrees of freedom:
\begin{align}
V_\mrm{1D}(|z_i-z_a|) = \iint d \bm{\rho}_i d \bm{\rho}_a
|\psi_0(\rho_i,\rho_a)|^2 V(|\mbf{r}_i -\mbf{r}_a|)
\end{align}
At sufficiently large distances the effective 1D interaction has similar power dependence as in 3D
\begin{align}
\label{V1Da}
V_\mrm{1D}(|z|) = - \frac{\alpha e^2}{2 z^4}, \quad |z| \gg l_\perp,
\end{align}
where $l_\perp = \hbar /(\mu \omega_{\perp})$. Summarizing, in our 1D calculations we apply the Hamiltonian \eqref{H1D} with the approximation \eqref{V1Da}, that are valid for $|z_i-z_a| \gg R_\mrm{1D} \equiv \max(R_\perp,l_\perp)$.

Finally in the case of equal longitudinal trapping frequencies: $\omega_i = \omega_a = \omega $, the relative and the COM degrees of freedom can be separated, and the dynamics is described by the Hamiltonian of the relative motion
\begin{align}
\label{Hrel1}
H_\mrm{rel1D} = \frac{p^2}{2 \mu} + \frac{1}{2} \mu \omega^2 (z - d)^2 - \frac{\alpha e^2}{2 z^4},
\end{align}
where $z=z_i-z_a$ and $p=p_i-p_a$. In the next Section we shall solve the problem in the different regimes described before, and investigate the peculiar phenomena that arise from the interplay between the trapping and the interaction potentials involving the two particles.

\subsection{Quantum-defect theory in 1D}
\label{Sec:QuantDef1D}

The quantum-defect treatment of the short-range interactions can be extended to the 1D dynamics, described by the Hamiltonian \eqref{H1D}. In 1D the asymptotic behavior of the relative wave function at short distances is governed by
\begin{align}
\label{Sol2}
\Psi_\mrm{rel}^e(z,k) & \sim |z| \sin\left(R^\ast/|z| + \varphi_e(k)\right), & \quad z \ll \sqrt{R^{\ast}/k},\\
\label{Sol3}
\Psi_\mrm{rel}^o(z,k) & \sim z \sin\left(R^\ast/|z| + \varphi_o(k)\right), & \quad z \ll \sqrt{R^{\ast}/k},
\end{align}
where labels $e$ and $o$ refer to the even and odd solutions respectively.
%We note that the asymptotic formulas \eqref{Sol2}-\eqref{Sol3} are valid in the regime of applicability of the Hamiltonian \eqref{H1D}, i.e., for $z \gg R_\mrm{1D}$, where the system exhibits 1D behavior, and for $z \ll R_i, R_a$, where one can neglect the trapping potentials.
In our model of 1D dynamics we treat Eqs.~\eqref{Sol2}-\eqref{Sol3} as boundary conditions for $|z| \rightarrow 0$.

In section~\ref{Sec:Renormalization} we show that the phases $\varphi_e$ and $\varphi_o$ are uniquely determined by $l_\perp$ and $\varphi$, and that their values can be calculated by solving the scattering problem in the quasi-1D geometry. This requires that $l_\perp \gg R_0$, which allows to use the quantum-defect description of the short-range potential. As we discuss later, in contrast to 3D traps, in quasi-1D traps the values of $\varphi$ are in general different for scattering waves of different symmetry.

\subsection{Time-dependent problem}
\label{Sec:TimeDep}

The atom-ion collision by moving the trapping potential $d(t)$ is an intrinsically time-dependent problem which requires the integration of the time dependent Schr\"odinger equation for the given initial condition to predict the transition probabilities for the possible final states. In our approach to the dynamics we first calculate the correlation diagrams, showing the energy levels as a function of the trap separations, and on the basis of these diagrams we predict the possible scenario of the atom-ion collision. When the motion of the trap is sufficiently adiabatic, the evolution of the system proceeds along one of the energy curves, therefore the basis of adiabatic eigenstates, dependent parametrically on $d$, is particularly useful in the analysis of the collision process. Of course, the adiabaticity is usually broken in the vicinity of avoided crossings. In such cases, however, one can apply e.g. the Landau-Zener theory to calculate the probability of the adiabatic and diabatic passage through an avoided crossing.

In our approach to the atom-ion collisions we consider that initially the traps are well separated. In this limit, the asymptotic states of \eqref{H3D} are given by products of the harmonic oscillator states in the two traps. In the course of dynamics, the distance $d(t)$ decreases, particle interact  for some definite time, and finally they are again separated, and the final state evolves into some superposition of the harmonic oscillator states (see Fig.~\ref{Fig:ContrColl} for a schematic picture). For such a scheme, we are interested in predicting the final state for some particular realization of $d(t)$. We stress that in the intermediate phase, the system may evolve into atom-ion molecular complex, and such possibility is fully accounted for in our model.

We start from the time-dependent Schr\"odinger equation with the Hamiltonian \eqref{H3D}. We expand the time-dependent wave function in the basis of the energy-ordered adiabatic eigenstates $\Psi_n(\mbf{x}_1,\mbf{x}_2|d)$ of the Hamiltonian \eqref{H3D}
\begin{equation}
H(d) \Psi_n(\mbf{x}_\mrm{i},\mbf{x}_\mrm{a}|d) = E_n(d) \Psi_n(\mbf{x}_\mrm{i},\mbf{x}_\mrm{a}|d),
\end{equation}
where we explicitly point out its dependence on the distance $d$. Substituting the expansion
\begin{align}
\Psi(\mbf{x}_\mrm{i},\mbf{x}_\mrm{a},t) = \sum_n & c_n(t) \exp\left[-\frac{i}{\hbar}
\int_{0}^{t} \!\!\!d\tau \, E_n(d(\tau))
\right] \nonumber \\
& \times \Psi_n\left(\mbf{x}_\mrm{i},\mbf{x}_\mrm{a}|d(t)\right)
\end{align}
into the Schr\"odinger equation, we obtain a set of coupled differential equations that govern the dynamics of the expansion coefficients
$c_n$:
\begin{align}
\dot{c}_n = - \dot{d} \sum_{m \neq n} & c_m(t)
 \exp\left[\frac{i}{\hbar}
\int_{0}^{t} \!\!\!d\tau \, \left(E_n(d(\tau))-E_m(d(\tau))\right)
\right] \nonumber \\
\label{cn}
& \times \langle \Psi_n(d)|{\tst \frac{\partial}{\partial d}} |\Psi_m(d) \rangle,
\end{align}
In the case of fully adiabatic evolution the coefficients $c_n(t) $ remain constant, and the evolution of the system proceeds along the adiabatic energy curves. From Eqs.~\eqref{cn} one can derive the condition for the adiabaticity of the transfer process. Adiabaticity requires that $\dot{d}$ multiplied by the nonadiabatic coupling $\langle \Psi_n(d)|{\tst \frac{\partial}{\partial d}} |\Psi_m(d)\rangle$ be much smaller than the frequency of the oscillating factor in the exponential of \eqref{cn}. In this case the oscillating factor effectively cancels out the contribution due to changes of $d$. The adiabaticity condition can be written as
$\hbar \dot{d} \langle \Psi_n(d)|\frac{\partial H}{\partial d} |\Psi_m(d) \rangle \ll (E_n - E_m)^2 \forall_{m,n}$ \cite{Tannor}. At trap separations where atom and ion can be approximately described by harmonic oscillator states, one can easily estimate the nonadiabatic couplings $\langle \Psi_n(d)| \frac{\partial H}{\partial d} |\Psi_m(d)\rangle$, which gives the following constraint on the adiabatic changes of the trap separation:
\begin{equation}
\frac{\dot{d}}{l_k} \ll \frac{(E_n - E_m)^2}{\hbar^2 \omega_k}, \quad k=i,a,
\end{equation}
where $k$ stands for $i$ ($a$) when the ion trap (atom trap) is moved. We stress that the latter condition is valid for transitions between different vibrational states, that may occur during the transfer of the atom or of the ion. In the case of avoided crossings between vibrational and molecular states, the adiabaticity of the transfer is characterized by the condition that can be determined from the Landau-Zener theory.

\section{Results}
\label{Sec:Results}

We turn now to the analysis of the adiabatic eigenenergies and eigenstates as a function of the trap separation $d$, and of the trapping potential geometry. We start our analysis from the simplest case of quasi-1D traps, where the dynamics takes place effectively in 1D, while the assumptions of equal trapping frequencies allows us to consider COM and relative motions separately. Before discussing this problem, we first study the dependence of the 1D short-range phases $\varphi_e$ and $\varphi_o$ on $l_\perp$, $R^{\ast}$ and $\varphi$. We argue that in the considered range of parameters they are practically independent of the kinetic energy of the scattering particles, which is assumed in our quantum-defect-theory approach. The reader not interested in details of the derivation may start at Section~\ref{Sec:Rel1D}, where we analyze 1D relative motion eigenenergies and eigenstates assuming some particular values of $\varphi_e$ and $\varphi_o$.  In Section~\ref{Sec:Rel3D} we switch to 3D geometries, analyzing the system of two spherically symmetric traps with the same trapping frequency for atom and ion. Section~\ref{Sec:AvoidedCross} is devoted to the properties of adiabatic energy spectra in the vicinity of an avoided crossing. Using the semiclassical theory we calculate the level separation at the avoided crossing, and then applying Landau-Zener theory we investigate the conditions for the adiabatic and diabatic transfer, depending on the distance between the traps. Finally, in Section~\ref{Sec:Full1D} we address the most complicated case of different trapping frequencies for atom and ion, coupling COM and relative degrees of freedom. Because of the complexity of this problem, we limit our analysis only to 1D dynamics; we argue, however, that the observed behavior should be qualitatively valid also for the 3D system.

\subsection{Short-range phases in quasi-1D traps}
\label{Sec:Renormalization}

As a preliminary technical step in preparation for the calculations of adiabatic energy curves that will be reported in the following, the present subsection deals with calculating the short-range phases $\varphi_e$ and $\varphi_o$ in quasi-one-dimensional geometries.

To find the 1D short-range phases we solve the Schr\"odinger equation for the relative part of the Hamiltonian \eqref{H3D}, assuming the same transverse trapping frequency for atom and ion: $\omega_{\perp i} = \omega_{\perp a} = \omega_\perp$, and neglecting axial trapping frequencies,
which is valid for $z \ll R_i,R_a$. The latter condition requires $R_i,R_a \gg R_\mrm{1D}$, where $R_\mrm{1D} = \max(R_\perp,l_\perp)$ determines
the boundaries of the 1D regime. In this way we obtain the following equation
for the relative wave function $\Psi(\mbf{r})$
\begin{equation}
\label{Schr1}
\left(-\frac{\hbar^2}{2 \mu} \nabla^2 - \frac{\alpha e^2}{2r^4} + \frac{1}{2} \mu \omega_\perp^2 \rho^2 - E \right)  \Psi(\mbf{r}) = 0,
\end{equation}
In the case of interaction potentials at distances $R_\mrm{1D}$
much larger than typical kinetic energies in the trapping potential: $|V(R_\mrm{1D})| \gg \hbar \omega_i, \hbar \omega_a$, we can neglect the energy dependence of the short-range phases, and solve \eqref{Schr1} for $E = \hbar \omega_\perp$.

At small distances, $r \ll R_\mrm{1D}$, the atom-ion interaction
dominates over the transverse trapping potential, and the solution
of \eqref{Schr1} behaves according to \eqref{Sol1}. At large
distances, $z \gg R_\mrm{1D}$, the motion in the transverse
direction is frozen to its zero-point oscillations and in the
asymptotic regime the scattering wave function assumes the form
$\Psi(\mbf{r}) \rightarrow \psi_0(\rho) \Psi_{1D}(z)$ ($|z|
\rightarrow \infty$), where $\psi_0(\rho)$ is the ground-state wave
function of the 2D harmonic oscillator: $\psi_0(\rho) = e^{-
\rho^2/(2 l_\perp^2)}/\pi^{1/2}$, and $\Psi_{1D}(z)$ is a linear
combination of odd and even waves: 
\begin{equation}
\label{Asympt1D}
\Psi_{1D}(z) = c_e \Psi_e(z)+ c_o \Psi_o(z), 
\end{equation}
with $\Psi_e(z)$, $\Psi_g(z)$ given by \eqref{Sol2}
and \eqref{Sol3} respectively \footnote{1D solutions \eqref{Sol2}
and \eqref{Sol3} are valid at all distances for $k \rightarrow 0$
($E \rightarrow \hbar \omega_\perp$).}.

\begin{figure}
   \includegraphics[width=8.6cm,clip]{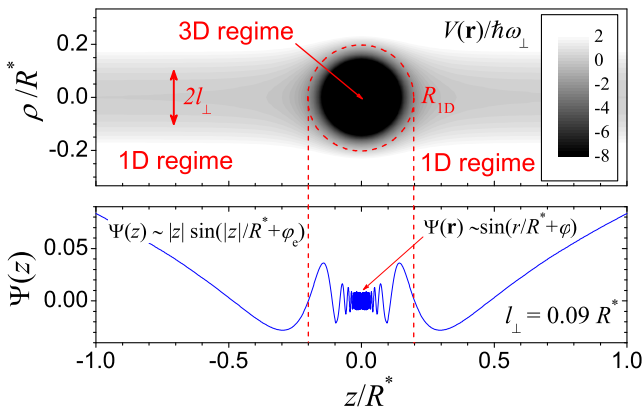}
     \caption{
     (Color online) Potential for the atom-ion relative motion (upper panel) and the relative wave function (lower panel) for the quasi-one dimensional system with $l_\perp = 0.09 R^{\ast}$ and without trapping in the axial direction.
     \label{Fig:Vrel1D}
   }
\end{figure}

Fig.~\ref{Fig:Vrel1D} shows the potential for the relative motion,
and the axial profile ($\rho=0$) of the relative wave function for
$l_\perp = 0.09 R^{\ast}$, which corresponds to a system of
$^{40}$Ca$^{+}$ and $^{87}$Rb in a trap with $\omega_\perp = 2 \pi
\times 1$MHz. In addition, the figure illustrates different length
scales present in the quasi-1D problem, and indicates the 1D ($|z|
\gg R_\mrm{1D}$) and 3D regimes ($r \ll R_\mrm{1D}$) in the behavior
of the wave function.

%%%%%%%%%%%%%%%%%% Figure 4 %%%%%%%%%%%%%%%%%%%%%%
%%%%%%%%%%%%%%%%%%%%%%%%%%%%%%%%%%%%%%%%%%%%%%%%%%%%%

\begin{figure}
   \includegraphics[width=8.6cm,clip]{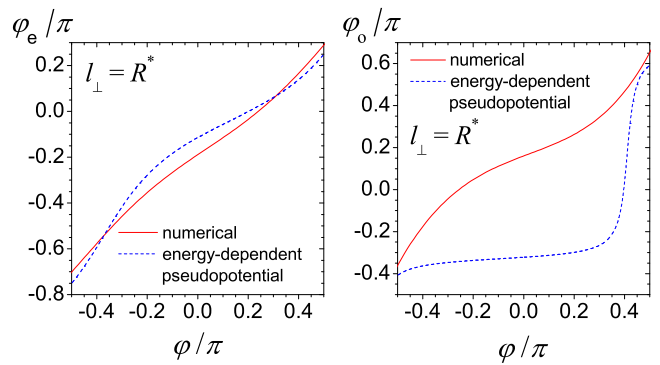}
     \caption{ 
     (Color online) Even and odd short-range phases $\varphi_e$ and $\varphi_o$ calculated for $l_\perp = R^{\ast}$. Numerical results (red solid lines) are compared with predictions of the model replacing $r^{-4}$ with the energy dependent pseudopotential.
     \label{Fig:phi1}
     }
\end{figure}

Fig.~\ref{Fig:phi1} shows an example of the dependence of
$\varphi_e$ and $\varphi_o$ on $\varphi$, for $l_\perp = R^{\ast}$. It compares the results of numerical calculations with predictions based on the pseudopotential method, that is dicsussed in appendix~\ref{App:Pseudo}. 
We observe that for even waves the agreement is fairly good, while for odd waves the agreement is poorer, which is probably due to the fact that the $p$-wave energy-dependent pseudopotential does not work already in the regime of $R^{\ast} \sim l_\perp$, or contributions of odd partial waves with $l > 1$ can be important. The pseudopotentials are expected to give accurate predictions for $R^{\ast} \ll l_\perp$.

%%%%%%%%%%%%%%%%%% Figure 5 %%%%%%%%%%%%%%%%%%%%%%
%%%%%%%%%%%%%%%%%%%%%%%%%%%%%%%%%%%%%%%%%%%%%%%%%%%%%

\begin{figure}
   \includegraphics[width=8.5cm,clip]{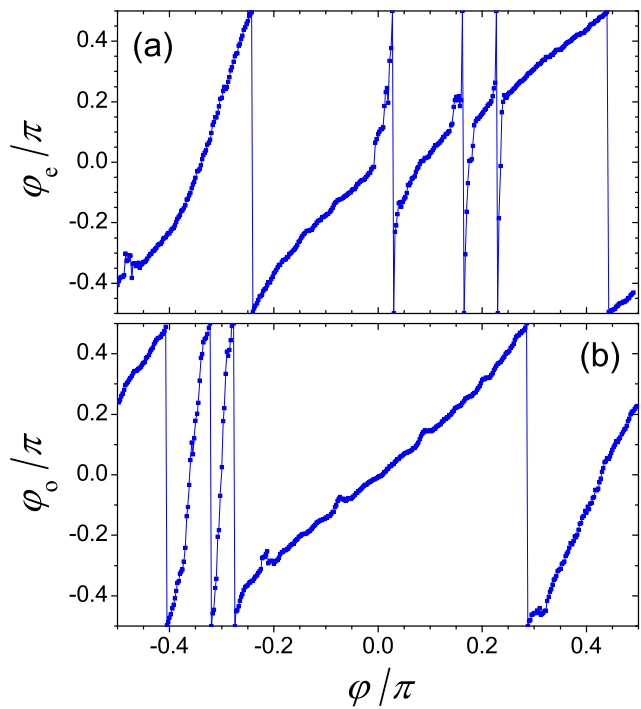}
     \caption{(Color online)
     \label{Fig:phi01}
     Even short-range phase $\varphi_e$ (upper panel) and odd short-range phase $\varphi_o$ (lower panel) calculated numerically for $l_\perp = 0.1 R^{\ast}$, versus 3D short-range phase $\varphi$.
     }
\end{figure}

\begin{figure}
   \includegraphics[width=8.5cm,clip]{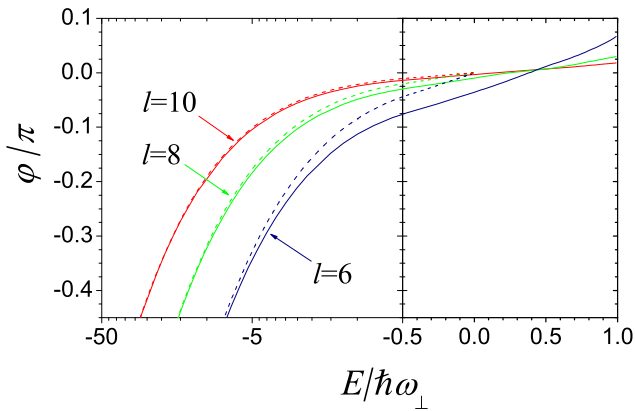}
     \caption{(Color online)
     \label{Fig:BoundSt}
     3D short-range phase $\varphi$ versus energy of bound states in the presence of the transverse confinement with $l_\perp = 0.1 R^{\ast}$. The bound states in the combined harmonic and $r^{-4}$ potentials (solid lines) are compared with the bound states of pure $r^{-4}$ interaction (dashed lines).
     }
\end{figure}

When $R^{\ast} \gg l_\perp$, the pseudopotential approach is not
applicable at all, and one has to resort to numerical calculations.
As an example, we present in Fig.~\ref{Fig:phi01} values for the
short-range phases for $l_\perp = 0.1 R^{\ast}$. We note the
presence of several resonances in the dependence of $\varphi_e$ and
$\varphi_o$. This behavior is related to the contribution of several
partial waves for $R^{\ast} \gg l_\perp$, leading to resonances when
the energy of a bound state in the combined harmonic and $r^{-4}$
potential becomes equal to the energy of the scattered wave. To
analyze this issue more carefully we have calculated the energies of
some particular bound states in in the combined harmonic and
$r^{-4}$ potential. Fig.~\ref{Fig:BoundSt} shows the relation
between the bound-state energy and the short-range phase for few
bound states that are responsible for the scattering resonances
close to $\phi=0$ for $\phi_e$ (cf. Fig.~\ref{Fig:phi01}). For
comparison we include the bound states of pure $r^{-4}$ potential
and we label them by the angular momentum $l$. We note that, due to
the limited resolution of our numerical calculations,
Fig.~\ref{Fig:phi01} does not allow to resolve the single resonances
close to $\phi=0$, however, one can observe the rapid changes of
$\phi_e$ in this region.

%%%%%%%%%%%%%%%%%% Figure 6 %%%%%%%%%%%%%%%%%%%%%%
%%%%%%%%%%%%%%%%%%%%%%%%%%%%%%%%%%%%%%%%%%%%%%%%%%%%%

%%%%%%%%%%%%%%%%%% Figure 7 %%%%%%%%%%%%%%%%%%%%%%
%%%%%%%%%%%%%%%%%%%%%%%%%%%%%%%%%%%%%%%%%%%%%%%%%%%%%

For the numerical calculations we have transformed \eqref{Schr1}
into cylindrical coordinates and we were solving 2D elliptic partial
differential equation using a finite-element method. To fix the
short-range phase we impose the boundary conditions on the
logarithmic derivative of $\Psi(r)$, $\partial_r \Psi(r)/\Psi(r)$,
at $r_\mrm{min}$, where $r_\mrm{min}= 0.09 R^\ast$ ($r_\mrm{min}=
0.022 R^\ast$) for $l_\perp = R^{\ast}$ ($l_\perp = 0.1 R^{\ast}$) .
At large distances we impose the Dirichlet boundary condition on the
rectangle with boundaries $|z| = z_\mrm{max}$, $\rho
=\rho_\mrm{max}$, assuming that the wave function has a Gaussian
transverse profile at $|z|=z_\mrm{max}$ and vanishes at
$\rho=\rho_\mrm{max}$. For $l_\perp = R^{\ast}$ we took
$z_\mrm{max}=6.3 R^\ast$, $\rho_\mrm{max}=3 R^\ast$, while for
$l_\perp = 0.1 R^{\ast}$ we have used $z_\mrm{max}=R^\ast$,
$\rho_\mrm{max}=0.4 R^\ast$. To determine the values of
$\varphi_e$ and $\varphi_o$, we fit at large distances ($|z| \gg R_\mrm{1D}$) the symmetric and antisymmetric solutions of \eqref{Schr1} for $E=\hbar \omega_\perp$, to the asymptotic formula \eqref{Asympt1D}.

\subsection{ Relative motion in a 1D system: adiabatic eigenenergies and eigenstates}
\label{Sec:Rel1D}

\begin{figure}
   \includegraphics[width=8.5cm,clip]{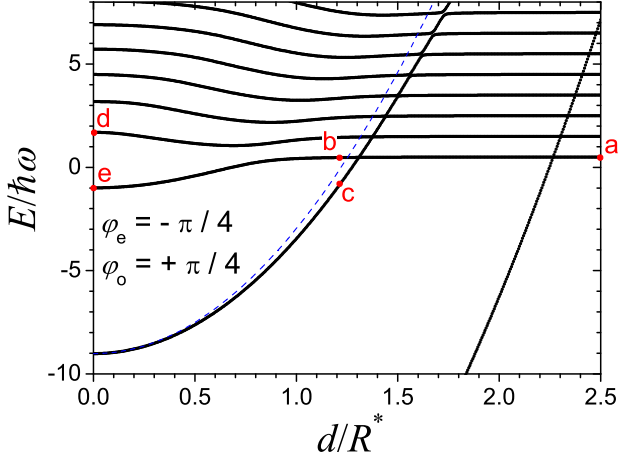}
     \caption{(Color online)
     Energy spectrum of the relative motion for an atom and an ion in 1D traps versus the distance $d$ between traps, calculated for $\varphi_e = - \pi/4$, $\varphi_o = \pi/4$ and  $R^{\ast} = 3.48 l$ (see text for details). The dashed curve shows the line $\frac{1}{2} \mu \omega^2 d^2$, giving the approximate shift of the bound state in the trapping potential (see section~\ref{Sec:Rel3D} for derivation).
    \label{Fig:E1DRel}
   }
\end{figure}

\begin{figure}
   \includegraphics[width=8.5cm,clip]{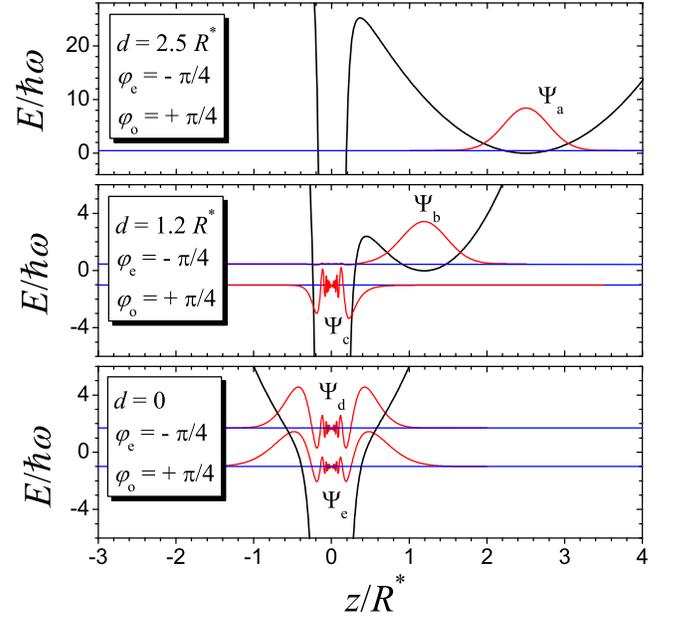}
     \caption{(Color online)
     Eigenstates of the relative motion for an atom and an ion in 1D traps at different separations $d$ between the traps, calculated for $\varphi_e = - \pi/4$, $\varphi_o = \pi/4$ and $R^{\ast} = 3.48 l$. Presented eigenstates correspond to labeled points in Fig.~\ref{Fig:E1DRel}. The horizontal lines present the corresponding eigenenergies, and the thick black line shows the potential energy.
     \label{Fig:1Dst}
   }
\end{figure}

In this section we consider the 1D motion of an atom and an ion, assuming $\omega_i = \omega_a$. In this case we can focus only on the relative motion
described by the Hamiltonian \eqref{Hrel1}, with the boundary conditions at $z \rightarrow 0$, stated by \eqref{Sol2} and \eqref{Sol3}. To find eigenenergies and eigenfunctions for arbitrary value of $d$ we diagonalize the Hamiltonian \eqref{Hrel1} numerically in the basis of its eigenstates for $d=0$, that are found by numerical integration of the 1D Schr\"odinger equation. In the basis we include the lowest 50 odd and even eigenfunctions, which is sufficient to perform the diagonalization for $d/R^\ast \lesssim 2.5$. Fig.~\ref{Fig:E1DRel} present the adiabatic energy spectrum for some example parameters: $\varphi_e = - \pi/4$, $\varphi_o = \pi/4$ and $R^{\ast} = 3.48 l$. The latter value corresponds to the system of $^{40}$Ca$^{+}$ and $^{87}$Rb in the trap with $\omega = 2 \pi \times 100$kHz, while the particular choice of the short-range phases $\varphi_e$ and $\varphi_o$ is explained later in this section. Points with labels correspond to the wave functions shown in Fig.~\ref{Fig:1Dst} presented together with their eigenenergies and potential energy curves. 

The three panels shown in Fig.~\ref{Fig:1Dst} illustrate three different regimes, where the system exhibits qualitatively different behavior. In the first regime, represented by the eigenstate $\Psi_a$ and realized at large distances between traps, $d \gg R_\mrm{rel}$ with $R_\mrm{rel}$ defined in \eqref{Rrel}, the main effect of the atom-ion interaction is the distortion of the trapping potential, and in this limit the Hamiltonian can be diagonalized analytically in the model of two coupled harmonic oscillators (see Section \ref{Sec:Full1D} for more details). The value of the short-range phase is not important for this model. At large separations the eigenstate $\Psi_a$ is typically only weakly perturbed with respect to the eigenstate of the harmonic oscillator.

In the second regime, represented by the eigenstates $\Psi_b$ and $\Psi_c$, and realized at distances $d \sim R_\mrm{rel}$, the system exhibits resonances between vibrational ($\Psi_b$) and molecular ($\Psi_c$) states, manifesting themselves as avoided crossings in the adiabatic energy spectrum.
The resonances appear when the energies of the two eigenstates become equal, and the energy splitting at the avoided crossing is proportional to the tunneling rate through the potential barrier separating the two regions of the potential. In section~\ref{Sec:AvoidedCross} we calculate this splitting in the framework of a semiclassical approximation, and describe the dynamics in the vicinity of the avoided crossing. The value of the short-range phase determines
the energy of the molecular states, but it is practically not important for the atom-ion vibrational states. Since the point of the avoided crossing depends on the short-range phases, the controlled collisions can provide some information on the short-range interaction potential.

Finally, at distances $d \ll R_\mrm{rel}$ the barrier separating the external trap from the well given by the atom-ion attraction disappears and all eigenstates have some short-range component behaving at short distances according to Eqs.~\eqref{Sol2} and \eqref{Sol3}. Hence, in this regime all eigenstates depend on the value of the short-range phases. The molecular states with energies below the dissociation threshold ($E=0$) are mainly localized in the well of the atom-ion attractive potential, however for the tight traps considered here, they can be strongly affected by the external trapping (e.g. state $\Psi_e$). The other type of states, with energies $E>0$, are analogs of the vibrational states and they are localized mainly at distances where the external trapping potential dominates
(e.g. state $\Psi_d$).

%%%%%%%%%%%%%%%%%% Figures 8 & 9 %%%%%%%%%%%%%%%%%%%%%%
%%%%%%%%%%%%%%%%%%%%%%%%%%%%%%%%%%%%%%%%%%%%%%%%%%%%%

\begin{figure}
   \includegraphics[width=8.5cm,clip]{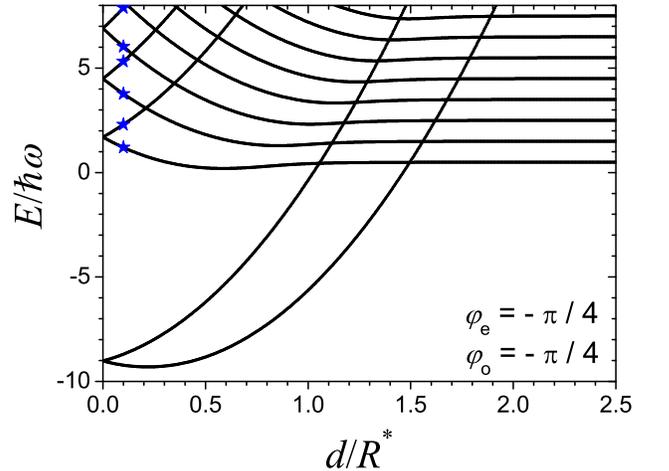}
     \caption{(Color online)
     Energy spectrum of the relative motion for trapped atom and ion in 1D traps versus distance $d$ between traps, calculated for $\varphi_e = \varphi_o = - \pi/4$ and $R^{\ast} = 3.48 l$ (see text for details).
    \label{Fig:E1DRel_sp}
   }
\end{figure}

\begin{figure}
   \includegraphics[width=8.5cm,clip]{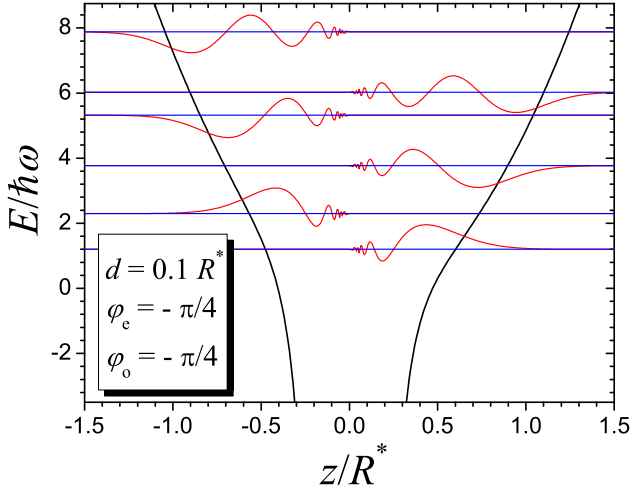}
     \caption{(Color online)
     Eigenstates of the relative motion for trapped atom and ion in 1D at trap separation $d=0.1 R^\ast$, calculated for $\varphi_e = \varphi_o = - \pi/4$ and $R^{\ast} = 3.48 l$. The horizontal lines present the corresponding eigenenergies, and the thick black line shows the potential energy.
     \label{Fig:1Dst_sp}
   }
\end{figure}

So far we have discussed the properties of the energy spectrum for some particular choice of $\varphi_e$ and $\varphi_o$. It turns out that a qualitatively similar behavior can be observed in all the systems with $|\varphi_e - \varphi_o| = \pi/2$. In the general case, however, the adiabatic energy spectrum has a slightly more complicated structure, as it is illustrated in Fig.~\ref{Fig:E1DRel_sp}, showing the eigenenergies for $\varphi_e = \varphi_o = - \pi/4$. We note that by going from $d = 0$ to positive $d \neq 0$ all the energies split into two branches. To understand the nature of this splitting in Fig.~\ref{Fig:1Dst_sp} we present the wave functions for $d=0.1 R^\ast$, with the corresponding eigenvalues marked with stars in Fig.~\ref{Fig:E1DRel_sp}. We observe that the branches represent the eigenstates localized on the left-hand side and right-hand side of the point $z=0$. The left-localized eigenstates correspond to the rising branches, because they are mainly localized in the region of strong atom-ion attraction. We note that this behavior is similar to the properties of eigenstates in a double-well potential, where appropriate wave functions are constructed by taking symmetric and antisymmetric combination of the states in two wells. In the particular case of $\varphi_e = \varphi_o$, the symmetric and antisymmetric combination leads to the states localized on the positive and negative $z$ semi axes, respectively.

%%%%%%%%%%%%%%%%%% Figure 10 & 11 %%%%%%%%%%%%%%%%%%%%%%
%%%%%%%%%%%%%%%%%%%%%%%%%%%%%%%%%%%%%%%%%%%%%%%%%%%%%

\subsection{Relative motion in 3D for spherical $\omega_i = \omega_a$ traps: adiabatic eigenenergies and eigenstates}
\label{Sec:Rel3D}

In this section we extend our analysis to 3D, and consider spherically symmetric traps for the atom and the ion with the same trapping frequencies: $\omega_i = \omega_a$. In this case the COM and relative motion can be separated, and in the following we focus only on the relative motion governed by the Hamiltonian \eqref{Hrel}.

We have diagonalized the Hamiltonian \eqref{Hrel} taking different values of the short-range phase $\varphi$, and we have observed that the adiabatic energy spectra exhibit qualitatively the same features as in 1D. In the numerical calculations we first calculated the eigenstates for $d=0$, by
solving the radial Schr\"{o}dinger equation, for angular momenta $l \leq 75$ and for energies $E \leq 75 \hbar \omega$. These states were used as a basis in the numerical diagonalization of the Hamiltonian \eqref{Hrel}.

\begin{figure}
   \includegraphics[width=8.5cm,clip]{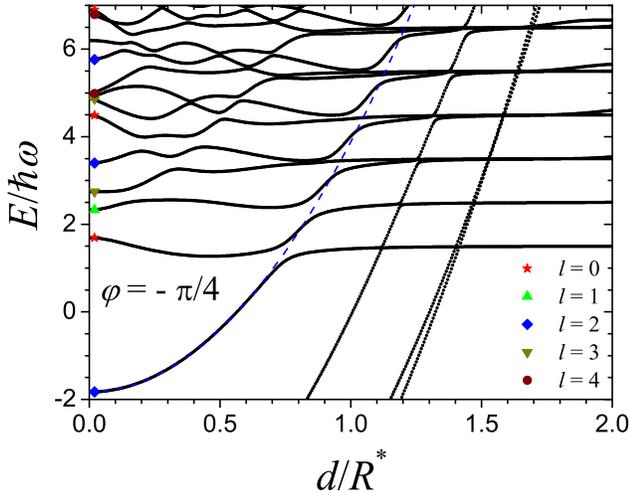}
     \caption{(Color online)
     Energy spectrum of the relative motion for an atom and an ion confined in spherically symmetric traps versus the distance $d$ between traps, calculated for $\varphi = - \pi/4$, and  $R^{\ast} = 3.48 l_r$ (see text for details). The blue dashed line shows the shift of the bound states with $d$, given approximately by $\frac{1}{2} \mu \omega^2 d^2$.
     \label{Fig:spectrum3D}
     }
\end{figure}

A sample adiabatic energy spectrum is shown in Fig.~\ref{Fig:spectrum3D},
presenting the results for $\varphi = -\pi/4$ and $R^{\ast} = 3.48 l$. At $d=0$ the angular momentum $l$ is a good quantum number and the states have definite angular symmetry, which is depicted by the appropriate symbols in
Fig.~\ref{Fig:spectrum3D}. For nonzero $d$, angular momentum is not conserved, hence in the dynamics the state can change its angular symmetry. 
We note that the energies of the bound states shift to a good approximation according to $\mu \omega^2 d^2 /2$. This behavior can be explained by noting that bound states $|\Psi_\mrm{mol}\rangle$ are concentrated around $r=0$, and  $\langle \Psi_\mrm{mol}(d)| H_\mrm{rel}(d) |\Psi_\mrm{mol}(d)\rangle \approx E_\mrm{b}+\frac{1}{2} \mu \omega^2 d^2$, where $E_{b}$ is the binding energy at $d=0$.

%%%%%%%%%%%%%%%%%% Figure 12 %%%%%%%%%%%%%%%%%%%%%%
%%%%%%%%%%%%%%%%%%%%%%%%%%%%%%%%%%%%%%%%%%%%%%%%%%%%%

\subsection{Avoided crossings: semiclassical analysis}
\label{Sec:AvoidedCross}

\begin{figure}
   \includegraphics[width=8cm,clip]{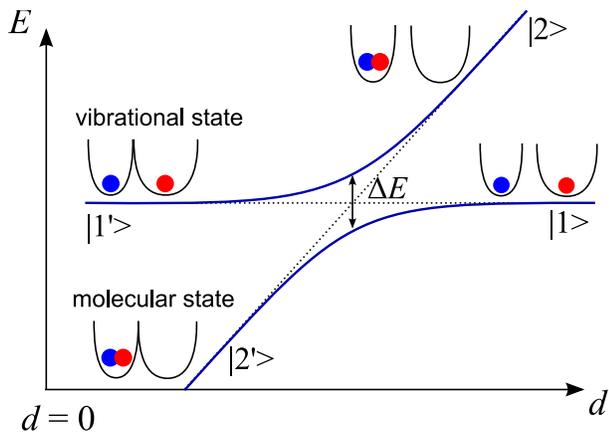}
     \caption{(Color online)
     Schematic drawing of the avoided crossing between vibrational and molecular states in the system of a trapped atom and an ion. An adiabatic change of the distance $d$ between traps induces a transition from the vibrational to the molecular states, while a diabatic process leaves the particles in the vibrational states.
     \label{Fig:AvoidedCross}
   }
\end{figure}

In the considered setup, level anticrossings reflect resonances between molecular and trap states. A simple picture of such an avoided crossing is shown in Fig.~\ref{Fig:AvoidedCross}. If one passes an avoided crossing from the direction of the asymptotic trap state, then for adiabatic evolution the system at small distances evolves into a molecular state. On the other hand, for diabatic passage, the particles remain in their traps, and the state basically does not change, apart from the modification due to the smaller trap separation. In addition, for fast changes of the trap positions, the particles can be excited to higher motional states. To describe quantitatively the dynamics in the vicinity of the avoided crossing one can
apply the Landau-Zener theory. Assuming that close to the avoided crossing the eigenenergies are linear in $d$, and that $d(t)$ varies linearly in time, the probability that the crossing is traversed diabatically is given by
\cite{Landau,Zener}
\begin{equation}
\label{ProbLZ}
p_{|1\rangle \rightarrow |1\rangle^{\prime}} = \exp \left(
- 2 \pi \frac{ |\langle \Psi_1|H|\Psi_2\rangle|^2}{\hbar |\dot{d}| |\partial E_{12}/\partial d|} \right),
\end{equation}
where the labels $|1\rangle$, $|2\rangle$ ($|1\rangle^{\prime}$, $|2\rangle^{\prime}$) respectively refer to the vibrational and molecular states before (after) passing the avoided crossing (cf.~Fig.~\ref{Fig:AvoidedCross}), and $E_{12}(d) = E_1(d) - E_2(d)$. For
$|\langle \Psi_1|H|\Psi_2\rangle|^2 \ll \hbar |\dot{d}| |\partial E_{12}/\partial d|$, the probability \eqref{ProbLZ} is close to unity, and the avoided crossing is passed diabatically. In the opposite case:
$|\langle \Psi_1|H|\Psi_2\rangle|^2 \gg \hbar |\dot{d}| |\partial E_{12}/\partial d|$, $p$ is small, and the avoided crossing is traversed adiabatically. The matrix element $\langle \Psi_1|H|\Psi_2\rangle$ can be related to the energy gap at the avoided crossing, and in this way we obtain the following constraint on the adiabaticity of the transfer close to the
avoided crossing:
\begin{equation}
\dot{d} \left| \frac{\partial E_1}{\partial d}
- \frac{\partial E_2}{\partial d}\right| \ll (E_1(d) - E_2(d))^2
\end{equation}

%%%%%%%%%%%%%%%%%% Figure 13 %%%%%%%%%%%%%%%%%%%%%%
%%%%%%%%%%%%%%%%%%%%%%%%%%%%%%%%%%%%%%%%%%%%%%%%%%%%%

The position of the avoided crossings is directly related to the energies of the bound states; thus, a measurement of the final state after controlled collisions, provides technique for spectroscopy of the trapped atom-ion complex. Since the avoided crossings become weaker as the separation between traps increases, this scheme allows to probe only excited molecular states, having sufficiently small binding energies, comparable to the energy scales of the trapping potentials.

In the case when the tunneling barrier is sufficiently large, the energy splitting at the avoided crossings can be estimated using a semiclassical approximation. We first focus on the relative motion in the 1D system, assuming $\omega_i = \omega_a = \omega$. Using the standard WKB method, one can derive the following result for the energy splitting $\Delta E$ \cite{Landau}
\begin{equation}
\label{WKB1D}
\Delta E = E \Psi_\mrm{mol}(x_1) \Psi_\mrm{vib}(x_2) \sqrt{v(x_1) v(x_2)} T
e^{-W/\hbar}.
\end{equation}
Here $x_1$, $x_2$ are some arbitrary points located in the classically forbidden region close to the classical turning points, $W(x_1,x_2,E) = \int_{x_1}^{x_2} dx \sqrt{2 m (V(x)-E)}$ is the action along the classical trajectory from $x_1$ to $x_2$, $T=\frac{\partial W}{\partial E}$ is the tunneling time, $\Psi_\mrm{mol}(x_1)$, $\Psi_\mrm{vib}(x_2)$ denote, respectively, the molecular and vibrational states (cf. Fig.~\ref{Fig:AvoidedCross}) with the same eigenvalues $E$ (the resonance case), and $v(x) = \sqrt{2(V(x)-E)/m}$ is the velocity of a particle
with energy $-E$ in the inverted potential $-V(x)$. In principle the choice of $x_1$ and $x_2$ is arbitrary, but in our calculations we take $x_1$ and $x_2$ at fixed distance from the location of the molecular ($x=0$) and
vibrational states ($x=d$), close to the classical turning points. In this way at sufficiently large separations $\Psi_\mrm{mol}(x_1)$, $\Psi_\mrm{vib}(x_2)$ becomes independent of $d$.

We turn now to the case of relative motion in 3D traps. Similar to the analysis for 1D systems, the energy splitting at the avoided crossings can be calculated semiclassically by applying the  instanton technique. In our calculation we adopt the formulation based on the path decomposition expansion developed by Auerbach and Kivelson \cite{Auerbach}.
We obtain the following formula describing the level splitting:
\begin{equation}
\label{WKB3D}
\Delta E = E A \Psi_\mrm{mol}(\mbf{x}_1) \Psi_\mrm{vib}(x_2) \sqrt{v(\mbf{x}_1) v(\mbf{x}_2)} T e^{-W/\hbar}
\end{equation}
The meaning of all the quantities is the same as in the 1D case, the only difference is the prefactor $A$ which accounts for the fluctuations around the instanton path (see \cite{Auerbach} for details). In the case $\omega_i = \omega_a$ the instanton path that minimizes the classical action is simply a straight line connecting the trap centers.

\begin{figure}
   \includegraphics[width=8.5cm,clip]{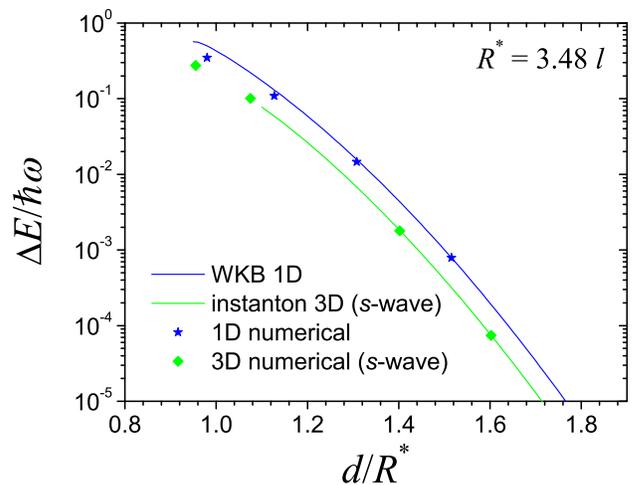}
     \caption{(Color online)
      Energy splitting at the avoided crossing between vibrational ground-state and the molecular state in 1D and 3D systems. Semiclassical results (solid lines) are compared to the numerical values extracted from the adiabatic energy spectra for different combinations of $\varphi_e$ and $\varphi_o$. 3D results present avoided crossings widths for $s$-wave bound states.
     \label{Fig:Prob1D}
     }
\end{figure}

Fig.~\ref{Fig:Prob1D} compares the semiclassical formulas Eqs.~\eqref{WKB1D} and \eqref{WKB3D} with the exact energy splitting determined from the numerical energy spectra, as those presented in Fig.~\ref{Fig:E1DRel}. The presented results correspond to the avoided crossings between the vibrational ground state molecular states. In 1D the energy splitting at the avoided crossing does not depend on the symmetry of the molecular state, while in 3D it does depend on its angular momentum $l$, and we present here only the case $l=0$. In the first approximation, the splitting depends on the short-range phases only through the critical trap separation at which the resonance occurs. The numerical data are obtained for different combinations of the short-range phases, for which the adiabatic spectra exhibit avoided crossings at different values of $d$. In our approach, instead of calculating $\Psi_\mrm{mol}(x_1)$ and $\Psi_\mrm{vib}(x_2)$, we fix the overall amplitude in \eqref{WKB1D} by fitting to the single point at the largest trap separation, where we expect the WKB approximation to be most accurate. The semiclassical curves stop at the distances where the potential barrier disappears. We observe that for the same ratio of $R^\ast$ to $l$ the splittings in 1D are larger than in 3D. Fig.~\ref{Fig:Prob1D}
shows the prediction of the semiclassical formula Eq.~\eqref{WKB1D}, and the energy splittings calculated numerically for molecular states with different $l$. We note that separations at the avoided crossing are largest for the spherically symmetric molecular states, with $l=0$.

%%%%%%%%%%%%%%%%%% Figure 14 %%%%%%%%%%%%%%%%%%%%%%
%%%%%%%%%%%%%%%%%%%%%%%%%%%%%%%%%%%%%%%%%%%%%%%%%%%%%

%%%%%%%%%%%%%%%%%% Figure 15 %%%%%%%%%%%%%%%%%%%%%%
\begin{figure}
   \includegraphics[width=8.5cm,clip]{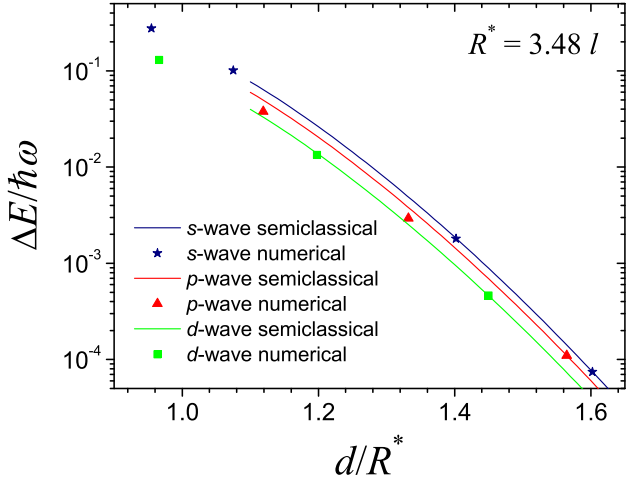}
     \caption{(Color online)
     Energy splitting at the avoided crossing between vibrational ground state and molecular states with different angular momenta, for a 3D spherically symmetric trap. Semiclassical results obtained by means of the instanton technique (solid lines) are compared to the numerical values extracted from the adiabatic energy spectra. Semiclassical calculations stop at the distance $d=1.1 R^{\ast}$ where the potential barrier disappears.
     \label{Fig:Prob3D}
     }
\end{figure}
%%%%%%%%%%%%%%%%%%%%%%%%%%%%%%%%%%%%%%%%%%%%%%%%%%%%%

The knowledge of the energy splitting $\Delta E$ can be used to calculate the probabilities of an adiabatic and diabatic passage of avoided crossings. Assuming that the avoided crossing is traversed at constant rate, we
apply the formula \eqref{ProbLZ} with $\partial E_1(d)/\partial d \approx 0$ and $\partial E_2(d)/\partial d = \langle \Psi_\mrm{mol}(d)| \partial H_\mrm{rel}(d)/ \partial d |\Psi_\mrm{mol}(d)\rangle \approx \mu \omega^2 d$
(c.f. Figs.~\ref{Fig:E1DRel} and \ref{Fig:spectrum3D}), which leads to
\begin{equation}
\label{ProbLZ1}
p_{|1\rangle \rightarrow |1\rangle^{\prime}} = \exp \left(
- \frac{\pi}{2} \frac{ (\Delta E)^2}{\hbar |\dot{d}| \mu \omega^2 d} \right),
\end{equation}
where we use the same notation as in Fig.~\ref{Fig:AvoidedCross}.
%We observe that the adiabatic passage of the avoided crossing requires
%$\hbar \dot{d} \mu \omega^2 d \ll (\Delta E)^2$.
Analyzing Figs.~\ref{Fig:Prob1D} and \ref{Fig:Prob3D} we can now estimate the rates $\dot{d}$ required for adiabatic and diabatic transitions. For instance in 3D, for the parameters of Fig.~\ref{Fig:Prob1D}, the diabatic transfer of particles across the avoided crossings up to distances $d \approx R^\ast$ ($\Delta E \lesssim 0.1 \hbar \omega$) can be realized by keeping
$\dot{d}/R^\ast \gg 0.001 \omega$ .

In summary, the analysis carried out here provides the basis for a description of the dynamics of various processes of interaction between an atom and an ion manipulated through external trapping potentials, and it gives a way to estimate quantitatively with simple analytical means the outcome of controlled interaction experiments in the different regimes described in Sect. II.

\subsection{Center of mass coupled to the relative motion: 1D analysis for $\omega_i \ne \omega_a$}
\label{Sec:Full1D}

In this section we consider the effects of coupling between COM and relative motions in 1D, that appear when the trapping frequencies for atom and ion are not equal.

\begin{figure}
   \includegraphics[width=8.6cm,clip]{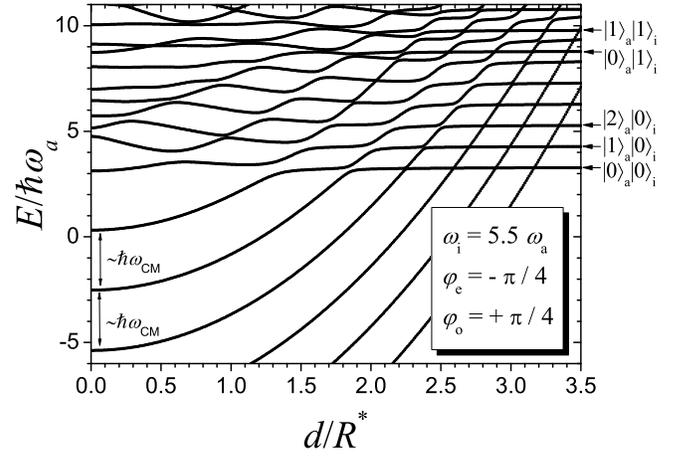}
     \caption{
     Energy spectrum for atom and ion confined in harmonic traps with $\omega_i = 5.5 \omega_a$, as a function of the distance $d$. Calculations are performed for $\varphi_e = - \pi/4$, $\varphi_o = \pi/4$ and $l_a = 0.9 R^{\ast}$ (see text for details).
     \label{Fig:SpectrFull}
     }
\end{figure}

Fig.~\ref{Fig:SpectrFull} shows the adiabatic levels as a function of trap separation for some example parameters: $\omega_i = 5.5 \omega_a$ and $l_a = 0.9 R^{\ast}$. This choice corresponds to the interaction of $^{40}$Ca$^{+}$ and $^{87}$Rb in the traps with $\omega_a = 2 \pi \times 10$kHz and
$\omega_i = 2 \pi \times 55$kHz. To obtain the adiabatic spectrum presented in Fig.~\ref{Fig:SpectrFull},
we performed the diagonalization of the Hamiltonian \eqref{H1D} in the product basis of the COM and relative motion eigenstates evaluated at $d=0$. In the calculations we consider all the states with a total energy $E \leq 460 \hbar \omega_a$, which leads to about 7100 states in the basis.

The arrows on the right-hand side of Fig.~\ref{Fig:SpectrFull}
indicate the asymptotic states for large separations, which can be
labeled by the number of phonons for the atom and ion trap,
respectively.  In comparison to the case where relative and COM
motions are decoupled, we observe the following new features: (i)
the molecular spectrum contains states with different numbers of
excitations in the COM degree of freedom. This can be observed at
$d=0$, when the molecular levels are with a good approximation
equally separated by $\hbar \omega_{CM}$ where $\omega_{CM}^2 = (m_a
\omega_a^2+ m_i \omega_i^2)/M$. \footnote{COM and relative motions
can be decoupled in the lowest order when the size of the molecule
is much smaller than the harmonic oscillator lengths $l_i$ and
$l_a$.}; (ii) the avoided crossings between molecular and
vibrational states are weaker for the states involving vibrational
excitations of the ion. This can be observed by comparing the
avoided crossings for the state $|0\rangle_a|1\rangle_i$ with the
avoided crossings for the neighboring states. This behavior can be
understood when we notice that for weaker atom trap (for $\omega_a <
\omega_i$), the particles have larger probability to tunnel when the
atom is excited.

%%%%%%%%%%%%%%%%%% Figure 16 %%%%%%%%%%%%%%%%%%%%%%
%%%%%%%%%%%%%%%%%%%%%%%%%%%%%%%%%%%%%%%%%%%%%%%%%%%%%

\subsection{Large distances between traps: two coupled oscillators approximation}
\label{Sec:Large}

Finally we turn to the limit of large distances: $d \gg R_{i},R_{a}$. Expansion of the interaction up to second-order terms in the distance $r$ leads to
\begin{align}
\label{Happr}
H_{1D} \simeq & \sum_{\nu = i,a} \left[ \frac{p_\nu^2}{2 m_\nu} +
\frac{1}{2} m_\nu \omega_\nu^2 (z_\nu-\bar{z}_\nu)^2 \right] \nonumber \\
& + 10 \frac{\alpha e^2}{d^6} (z_i-\bar{z}_i)(z_a-\bar{z}_a),
\end{align}
where $\bar{z}_\nu$ denotes the equilibrium position of the particles and
we neglect the second-order terms modifying the trapping frequencies.
The Hamiltonian (\ref{Happr}) describes the system of two coupled harmonic oscillators, which can be written in the form
\begin{equation}
\label{Hz}
H_{1D} = \hbar \omega_a a^\dagger a + \hbar \omega_i b^\dagger b +
\hbar \omega_c (a + a^\dagger)(b + b^\dagger),
\end{equation}
where we have introduced the usual annihilation and creation operators
\begin{align}
a & = \sqrt{\frac{m_i \omega_i}{2\hbar}}\left[z_i + i \frac{p_i}{m_i \omega_i}\right] \\
b & = \sqrt{\frac{m_a \omega_a}{2\hbar}}\left[z_a + i \frac{p_a}{m_a \omega_a}\right].
\end{align}
Here $\omega_c$ denotes the coupling frequency: $\hbar \omega_c = 10 E^\ast (R^\ast)^4  l_{a} l_{i} /d^6$. The validity of the considered model is
limited by the assumption that a stable equilibrium position exists,
which is fulfilled for $R_i,R_a \lesssim 0.57 d$, as it can be easily verified. Typical, maximal values of $\omega_c$ for atoms in optical potentials and ions in rf traps are of the order of $10$kHz.
In practice the model ceases to be valid already at weaker conditions, when the terms higher than the second order cannot be neglected.
In any case, this model is relevant for an important class of processes involving coherent transfer of quanta between the atom and the ion, of direct application in a quantum information processing context.

\section{Conclusions and outlook}
\label{Sec:Outlook}

In this paper we analyzed in detail the interaction between a single atom and a single ion guided by external trapping potentials. This kind of work is motivated by recently opened experimental possibilities within combined systems --~currently being built in several groups worldwide~-- where magneto-optical traps or optical lattices for neutral atoms coexist with electromagnetic traps for ions.

Tight confinement of single particles, associated with independent control of the respective confining potentials, allows for exploring different regimes of the two-body dynamics involving one atom and one ion at a time. At large distances, the interaction is dominated by the inverse quartic term arising from the attraction between the ion's charge and the electric dipole induced by it on the atomic electron wave function. A pseudopotential approximation, similar to that familiar from ultracold-atom collision theory, is not applicable in our case, as the characteristic range of the atom-ion molecular potential often exceeds the size of the tight traps experimentally available. To describe the interaction at short distances, smaller than that range, we employ quantum-defect theory, which allows to deal with different geometries from isotropic three-dimensional traps to very elongated quasi-one-dimensional ones.

A good description of adiabatic dynamics, involving processes where the traps are moved toward or across each other at a rate much slower than the trapping frequencies, can be obtained from quasistatic eigenenergy curves, which we calculate for various trapping configurations based on the methods outlined above. A remarkable feature displayed by the system is the presence of resonances between molecular-ion bound states and motional excitations within the trap. These trap-induced resonances are similar in nature to Feshbach resonances driven by external fields, and they could as well be spectroscopically probed in simple experiments where the interaction is controlled via the external guiding potentials.

In addition to the above aspects, an important motivation for the interest in systems of trapped atoms and ions and their trap-induced resonances resides in possible applications to quantum information processing. In this context, one can utilize controlled atom-ion interactions to effect coherent transfer of qubits, thereby creating interfaces between atoms and ions. By storing quantum information in internal atomic states, and performing gate operations with ions, one would combine the advantages of both: (i) long decoherence times for neutral atoms and (ii) short gate-operation times for charged particles due to the relatively strong interactions.
%In this case one can exploit the phases acquired by particles, during the controlled collision, provided that the motion of the trap is sufficiently adiabatic. By choosing appropriate interaction times, one can realize for instance a SWAP gate, that allows for exchange of qubit states between atoms and ions \cite{SympCool}.
In the present work we have focused on the motional degrees of freedom, which can serve as auxiliary degrees of freedom for quantum gates involving internal-state qubits, provided an appropriate coupling mechanism between internal and external degrees of freedom is employed -- e.g., sideband excitation via a laser.

Another potential application of our results is cooling of the
atomic motion. In typical neutral-atom quantum computation schemes
with qubits stored in internal states of atoms trapped in
optical-lattice sites, motional excitations constitute a serious
source of errors, and the atomic motion needs to be cooled in a
state-insensitive manner between computational steps in order to
avoid qubit decoherence. Since the long-range part of the atom-ion
interaction is not sensitive to the internal state \footnote{At
least for the internal states with the total angular momentum
$J=1/2$.}, our setup can be applied for sympathetic cooling of
atoms, through the exchange of energy with laser cooled ions
\cite{SympCool}.

Beside these applications in the context of quantum information processing, our work opens broader perspectives for the study of new interesting collisional physics in a physical situation never explored before.
In principle, our scheme allows for production of ultracold trapped atom-ion molecules (molecular complexes), when the trapping potentials are lowered adiabatically at the stage when the particles remain close to each other. In this way our method can be regarded as a way to perform cold chemical reactions, where the final state of the molecule can be well controlled.  Beyond sufficiently large separation between traps, however, the survival probability of such molecules in the final state is negligible. Indeed, in the calculations presented here, we have only regarded processes leading to a final state in which the atom and the ion are still separated, and no molecule has been formed as an outcome of the interaction. In other words, the Hamiltonian describing our single-channel model does not include the possibility of transitions to other Born-Oppenheimer curves. While this is a justified assumption for the dynamics considered here, it is in a sense a limitation of our current approach. Investigation of molecular formation in traps beyond this approximation is certainly among the interesting developments that can arise from our work. Its long-term motivation is really to open, beyond the present examples, a new paradigm for cold collision physics, which can be described as the mechanical control of single-particle chemical reactions.

\begin{acknowledgments}
The authors thank J. Denschlag, P. Schmidt, R. Stock, and A. Simoni for helpful discussions. This work was supported by the Austrian Science
Foundation (FWF), the European Union projects OLAQUI
(FP6-013501-OLAQUI), CONQUEST (MRTN-CT-2003-505089), the SCALA
network (IST-15714), the Institute for Quantum Information GmbH, the
EU Marie Curie Outgoing International Fellowship QOQIP, and the
National Science Foundation through a grant for the Institute for
Theoretical Atomic, Molecular and Optical Physics at Harvard
University and Smithsonian Astrophysical Observatory.
\end{acknowledgments}

\begin{appendix}

\section{Derivation of the effective Hamiltonian \eqref{H3D}}
\label{App:Micro}

While the Hamiltonian \eqref{H3D} is intuitively obvious, we find it nonetheless worthwhile to summarize the microscopic derivation in an adiabatic approximation, and to discuss its validity. For simplicity we consider an atom with a single outer-shell electron (alkali atom) and an ion with single positive charge $+ e$. In addition we do not consider the internal structure of the ion. The total Hamiltonian can be written as
\begin{equation}
H = \frac{\mbf{p}_1^2}{2 m_1} + \frac{\mbf{p}_2^2}{2 m_2} + H_\mrm{e} + H_\mrm{las}+ H_\mrm{rf}.
\end{equation}
Here the labels $1$ and $2$ correspond, respectively, to atomic nucleus and ion, and $H_\mrm{e}$ is the Hamiltonian of the electron, which includes the Coulomb interactions,
\begin{equation}
H_\mrm{e} = \frac{\mbf{p}_\mrm{e}^2}{2 m_\mrm{e}} - \frac{e^2}{|\mbf{x}_e - \mbf{x}_1|} - \frac{e^2}{|\mbf{x}_e - \mbf{x}_2|} + \frac{e^2}{|\mbf{x}_2 - \mbf{x}_1|}.
\end{equation}
Here, for simplicity, we omit the contributions of the core regions for both the atom and ion. In this way our model refers in fact to H$_2^+$ molecule. For alkali atoms and alkaline-earth ions one should treat the complete structure of core regions, however, our approach can be readily generalized to this more complicated case.

The term $H_\mrm{las}$ describes the interaction of the atom with a laser beam creating an optical potential, written in the electric dipole representation
\begin{equation}
H_\mrm{las} = - \mbf{d} \mbf{E}_\perp(\mbf{x}_1,t),
\end{equation}
where $\mbf{d} = e (\mbf{x}_1 - \mbf{x}_e)$ is the dipole moment of the atom, and for simplicity we neglected the influence of the laser on the ion, which has typically a different electronic structure that the atom. In addition we have applied the long-wavelength approximation, neglecting changes of the electric field on the scale of the atom. In the case of the optical lattice the transverse part of the electric field $\mbf{E}_\perp$ can be assumed to have the form of a standing wave
\begin{equation}
\mbf{E}_\perp(\mbf{x},t) = \mbf{E}_0 \cos (\omega_L t) \left[
\cos (k_L x+\phi_x) + ( x \rightarrow y,z) \right],
\end{equation}
where for simplicity we have assumed the same amplitude $\mbf{E}_0$, the same wave vector $k_L$, and the same frequency of the laser light $\omega_L$ for all three laser beams creating the optical lattice potential. The abbreviation $( x \rightarrow y,z)$ denotes sum of the terms with $x$ replaced by $y$ and $z$, and $\varphi_k$ for $k=x,y,z$ is the phase factor characterizing the standing wave. Finally, $H_\mrm{rf}$ is the electric potential creating the rf-trap
\begin{equation}
H_\mrm{rf} = e \Phi(\mbf{x}_1,t) + e \Phi(\mbf{x}_2,t) - e \Phi(\mbf{x}_\mrm{e},t),
\end{equation}
where $\Phi(\mbf{x}_1,t)$ is the time-dependent electric field of the rf trap
\begin{align}
\Phi(\mbf{x},t) = & \frac{1}{2} (u_x x^2 + u_y y^2 + u_z z^2) \nonumber \\
& + \frac{1}{2} (v_x x^2 + v_y y^2 + v_z z^2) \cos \omega_\mrm{rf} t.
\end{align}
Here, $\omega_\mrm{rf}$ is the frequency of the time-dependent part of the electric potential, and $u_k$, $v_k$ ($k=x,y,z$) are amplitudes depending on the geometry of the trap \cite{Leibfried}. The electric field at every instant of time has to fulfill the Laplace equation: $\Delta \Phi =0$, hence, the coefficients $u_k$, $v_k$ are subject to the following conditions:
$u_x+u_y+u_z =0$, $v_x+v_y+v_z=0$.

Below we indicate the basic steps of the derivation.

{\em Expansion in the basis of Born-Oppenheimer wave functions for electron motion}. We start from generating a complete set of electronic wave functions, parametrized by the positions of the atomic core and of the ion
\begin{equation}
H_e \Phi_n(\mbf{x}_e|\mbf{x}_1,\mbf{x}_2) = E_n(|\mbf{x}_2 - \mbf{x}_1|) \Phi_n(\mbf{x}_e|\mbf{x}_1,\mbf{x}_2)
\end{equation}
In this way the total wave function can be expanded in the basis of
Born-Oppenheimer electronic wave functions
\begin{equation}
\label{Exp}
\Psi(\mbf{x}_1,\mbf{x}_2,\mbf{x}_e,t) = \sum_{n} c_n(\mbf{x}_1,\mbf{x}_2,t) \Phi_n(\mbf{x}_e|\mbf{x}_1,\mbf{x}_2).
\end{equation}
Since the basis is complete, the expansion of the wave function does not involve any approximations.

{\em Retaining in the expansion only the modes coupled by the
laser}. In the expansion \eqref{Exp} we keep only two modes coupled
by the laser light creating the optical lattice: the electronic
ground state $\Phi_g(\mbf{x}_e|\mbf{x}_1,\mbf{x}_2)$ and the
electronic excited state $\Phi_e(\mbf{x}_e|\mbf{x}_1,\mbf{x}_2)$. In
this way we utilize the Born-Oppenheimer approximation, treating the
electron motion in the adiabatic approximation. This assumes that
the time scale of the electron dynamics is much faster than the
dynamics of the atomic nucleus and of the ion, which is typically
fulfilled since the electron is much lighter than the other two
particles. The approximation of the two coupled channels can be
easily generalized to more, or even infinite number of channels,
since the other channels are weakly populated and we treat them
within the perturbation theory.

{\em Adiabatic elimination of the excited electronic state coupled through the laser: derivation of the optical trap potential}.
The expansion coefficients $c_g(\mbf{x}_1,\mbf{x}_2,t)$ and $c_e(\mbf{x}_1,\mbf{x}_2,t)$ fulfill the following set of coupled equations:
\begin{widetext}
\begin{align}
\label{dcg}
i \hbar \frac{\partial c_g(\mbf{x}_1,\mbf{x}_2,t)}{\partial t} & =  \left(
\frac{\mbf{p}_1^2}{2 m_1} + \frac{\mbf{p}_2^2}{2 m_2} + E_g(|\mbf{x}_2 - \mbf{x}_1|) + e \Phi(\mbf{x}_2,t) \right) c_g(\mbf{x}_1,\mbf{x}_2,t) -
\mbf{d}_\mrm{eg} \mbf{E}_\perp(\mbf{x}_1,t) c_e(\mbf{x}_1,\mbf{x}_2,t) \\
\label{dce}
i \hbar \frac{\partial c_e(\mbf{x}_1,\mbf{x}_2,t)}{\partial t} & =  \left(
\frac{\mbf{p}_1^2}{2 m_1} + \frac{\mbf{p}_2^2}{2 m_2} + E_e(|\mbf{x}_2 - \mbf{x}_1|) + e \Phi(\mbf{x}_2,t) \right) c_e(\mbf{x}_1,\mbf{x}_2,t) -
\mbf{d}_\mrm{eg} \mbf{E}_\perp(\mbf{x}_1,t) c_g(\mbf{x}_1,\mbf{x}_2,t),
\end{align}
\end{widetext}
where $\mbf{d}_\mrm{eg}(\mbf{x}_1,\mbf{x}_2) = \langle \Phi_e|\mbf{d}|\Phi_g \rangle$ is the dipole matrix element between the ground and excited electronic states, which in general depends on the position of atom and ion.
In the derivation of \eqref{dcg}-\eqref{dce} we have neglected the action of the rf field on the atom core and the electron, putting
\begin{eqnarray}
\langle \Phi_k | H_\mrm{rf} | \Phi_k \rangle & \approx & e \Phi(\mbf{x}_2,t), \quad k=e,g \\
\langle \Phi_e | H_\mrm{rf} | \Phi_g \rangle & \approx & 0.
\end{eqnarray}
We focus on the regime of far-detuned laser: $\Delta \gg \Omega_L$, where $\Delta$ denotes the detuning $\Delta (|\mbf{x}_2 - \mbf{x}_1|)= E_e(|\mbf{x}_2 - \mbf{x}_1|) -E_g(|\mbf{x}_2 - \mbf{x}_1|) - \omega_L$, and $\Omega_L = - \mbf{d}_\mrm{eg} \mbf{E}_0/\hbar$ is the the Rabi frequency. For such conditions, the excited state is only weakly populated and can be adiabatically eliminated. Additional simplification comes from the fact that the transitions between states $g$ and $e$, due to the laser light, occur on a time scale much shorter than the motion of atom and ion, and the dynamics of the atom and ion centers of masses can be decoupled from the internal dynamics, in full analogy to the Born-Oppenheimer approximation. Eliminating the excited state, in basically the same manner as in the standard derivation of the AC Stark shift, we obtain the following equation that governs the dynamics of the atom and ion center of masses:
\begin{align}
i \hbar \frac{\partial c_g(\mbf{x}_1,\mbf{x}_2,t)}{\partial t}  =  \bigg( &
\frac{\mbf{p}_1^2}{2 m_1} + \frac{\mbf{p}_2^2}{2 m_2} + E_g(|\mbf{x}_2 - \mbf{x}_1|) + V_\mrm{opt}(\mbf{x}_1) \nonumber  \\
& + e \Phi(\mbf{x}_2,t) \bigg) c_g(\mbf{x}_1,\mbf{x}_2,t),
\end{align}
where
\begin{equation}
\label{Vopt}
V_\mrm{opt}(\mbf{x}) = - \frac{\hbar \Omega^2_L}{4 \Delta} \left( \cos (k_L x+\phi_x)^2 + ( x \rightarrow y,z) \right)
\end{equation}
is the effective potential due to the laser field. In the formula \eqref{Vopt}
we have neglected the position dependence of the detuning $\Delta (|\mbf{x}_2 - \mbf{x}_1|)$ and of the dipole matrix element $\mbf{d}_\mrm{eg}(\mbf{x}_1,\mbf{x}_2)$, assuming that they are modified only at very short distances between the particles. This can be justified, since in the atomic collisions, the probability of finding the particles at short distances, comparable to the range of chemical binding forces, is typically very small.

{\em Time averaging over fast oscillations of the rf field: derivation of the Paul trapping potential}. We replace the time-dependent rf field by an effective, adiabatic potential. In this way we neglect the fast micromotion of the ion on the time scale of $\omega_\mrm{rf}$ \cite{Leibfried}. Finally we obtain the following equation that describes the dynamics of the wave function $c_g(\mbf{x}_1,\mbf{x}_2,t)$ dependent on the atom and ion positions
\begin{align}
i \hbar \frac{\partial c_g}{\partial t}  =  \bigg( &
\frac{\mbf{p}_1^2}{2 m_1} + \frac{\mbf{p}_2^2}{2 m_2} + E_g(|\mbf{x}_2 - \mbf{x}_1|) + V_\mrm{opt}(\mbf{x}_1) \nonumber  \\
& V_\mrm{rf}(\mbf{x}_2) \bigg) c_g(\mbf{x}_1,\mbf{x}_2,t).
\end{align}
Here $V_\mrm{rf}$ denotes the effective potential
\begin{equation}
V_\mrm{rf}(\mbf{x}) = \frac{1}{2} m_2 \left(
\omega_x x^2 +\omega_y y^2 + \omega_z z^2 \right),
\end{equation}
where $\omega_k = (a_k +q_k^2/2)^{1/2} \omega_\mrm{rf} /2$, $a_k = 4 e u_k/(m_2 \omega_\mrm{rf}^2)$ and $q_k = 2 e v_k/(m_2 \omega_\mrm{rf}^2)$  for $k=x,y,z$. The interaction between atom and ion is given by $E_g(|\mbf{x}_2 - \mbf{x}_1|)$ -- the ground-state energy of the electron motion, which at large distances behaves as $E_g(r) \sim -\alpha e^2/(2r^4)$.

\section{Determining of $\varphi_e$ and $\varphi_o$ from pseudopotentials}

\label{App:Pseudo}
In principle, the replacement of the $r^{-4}$ interaction by the pseudopotential is strictly valid when $R^{\ast} \ll l_{\perp}$. Nevertheless, it is possible to try to determine $\varphi_e$ and $\varphi_o$ for $R^{\ast} \sim l_{\ast}$ using the energy-dependent pseudopotentials.
To this end we replace the $r^{-4}$ interaction with \cite{Blume,Bolda}
\begin{equation}
V_s(\mbf{r}) = \frac{2 \pi \hbar^2 a(k)}{\mu} \delta(\mbf{r}) \frac{\partial}{\partial r} r
\end{equation}
for even scattering, and \cite{Idziaszek}
\begin{equation}\label{Vp}
V_{p}(\mbf{r}) = \frac{\pi \hbar^2 a_p(k)^3}{\mu}
\stackrel{\leftarrow}{\nabla}
\delta(\mbf{r}) \stackrel{\rightarrow}{\nabla}
r \frac{\partial^3}{\partial r^3} r^2,
\end{equation}
for odd scattering. Here, $a(k) = -\tan \delta_0(k)/k$ is the energy-dependent $s$-wave scattering length, $a_p$ is the $p$-wave scattering length: $a_p(k)^3 = - \tan \delta_1(k) /k^3$, the symbol $\stackrel{\leftarrow}{\nabla}$
($\stackrel{\rightarrow}{\nabla}$) denotes the gradient operator
that acts to the left (right) of the pseudopotential, and
$\delta_0(k)$, $\delta_1(k)$ are the $s$- and $p$-wave phase shifts, respectively. 

In 1D the even ($\psi_e$) and odd ($\psi_o$) scattering waves have the following asymptotic behavior: $\psi_e(z,k) \sim \sin (k |z| + \delta_e(k))$, and $\psi_o(z,k) \sim z/|z| \sin (k |z| + \delta_o(k))$ ($|z| \rightarrow \infty$), where $\delta_e(k)$ and $\delta_o(k)$ denote even and odd scattering phases, respectively. In analogy to the three-dimensional scattering theory, we define 1D even ($a_\mrm{1D}^\mrm{e}$) and 1D odd ($a_\mrm{1D}^\mrm{o}$) scattering lengths: $a_\mrm{1D}^\mrm{e,o} = \lim_{k \rightarrow 0} - (\tan \delta_{e,o}(k))/k$. From this definition it follows that for $k=0$ the even and odd scattering waves behaves as: $\psi_e(z,k=0)
\sim |z| - a_\mrm{1D}^\mrm{e}$ and 
$\psi_o(z,k=0) \sim z - a_\mrm{1D}^\mrm{o} z/|z|$ ($|z| \rightarrow \infty$). The latter result applied to Eqs.~\eqref{Sol2} and \eqref{Sol3} leads to  
\begin{align}
a_\mrm{1D}^\mrm{e}/R^\ast & = - \cot \varphi_\mrm{e} \\
a_\mrm{1D}^\mrm{o}/R^\ast & = - \cot \varphi_\mrm{o}
\end{align}
In the pseudopotential approximation one can solve the quasi-1D scattering problem exactly, and calculate values of the 1D scattering lengths
for even \cite{Bergeman,Idziaszek2} and odd \cite{Granger} waves
\begin{align}
\label{a1De}
a_\mrm{1D}^\mrm{e}(k) & = -\frac{l_\perp^2}{2 a_s(E)} - \frac{l_\perp}{2}
\zeta \left(\frac{1}{2},\frac{3}{2} - \frac{E}{2 \hbar \omega_\perp}\right), \\
\label{a1Do}
a_\mrm{1D}^\mrm{o}(k) & = \frac{l_\perp}{2} \left[
 \frac{l_\perp^3}{12 a_p(E)^3} - 
\zeta \left(-\frac{1}{2},\frac{3}{2} - \frac{E}{2 \hbar \omega_\perp}\right) \right]^{-1},
\end{align}
where $E = \hbar \omega_\perp + \hbar^2 k^2 /(2 \mu)$ and $\zeta(s,a)$ denotes the Hurwitz Zeta function: 
$\zeta(s,a) =  \sum_{k=0}^{\infty} (k+a)^{-s}$ 
\cite{Abramowitz}. Finally,  to relate the energy-dependent scattering lenghts $a_s(E)$ and $a_p(E)$ to $\varphi$, we apply the exact solutions of the Schr\"odinger equation for $r^{-4}$ potential, given by the Mathieu functions \cite{Vogt,OMalley,Spector}.

\end{appendix}

\end{document}